\documentclass[aps,twocolumn]{revtex4-1}
\usepackage{amsmath,amssymb}
\usepackage{slashed}
\usepackage{graphicx}
\usepackage{bm}
\usepackage[top=1.0in,bottom=1.0in,left=1.0in,right=1.0in]{geometry}
\usepackage[colorlinks,linkcolor=blue,citecolor=blue]{hyperref}

\newcommand{\be}{\begin{equation}}
\newcommand{\ee}{\end{equation}}
\newcommand{\bea}{\begin{eqnarray}}
\newcommand{\eea}{\end{eqnarray}}
\newcommand{\ba}{\begin{eqnarray}}
\newcommand{\ea}{\end{eqnarray}}

\begin{document}

\title{Meson structure on the light-front  III \\
The Hamiltonian, heavy quarkonia, spin and orbit mixing}

\author{Edward Shuryak}
\email{edward.shuryak@stonybrook.edu}
\affiliation{Center for Nuclear Theory, Department of Physics and Astronomy, Stony Brook University, Stony Brook, New York 11794--3800, USA}

\author{Ismail Zahed}
\email{ismail.zahed@stonybrook.edu}
\affiliation{Center for Nuclear Theory, Department of Physics and Astronomy, Stony Brook University, Stony Brook, New York 11794--3800, USA}

\begin{abstract}
This is the third paper on hadronic  light front wave functions (LFWFs). We derive
 a light front Hamiltonian from first principles,
using the key features of the QCD vacuum at low resolution. 
In the first approximation, it  gives  transverse oscillator  and longitudinal harmonic modes,
and yields the  correct Regge trajectories.
For heavy quarkonia, we compare its spectrum to that obtained from
the usual Schroedinger equation in the rest frame. We use the same approach
for light quarks, and investigate the role of confinement and chiral symmetry 
breaking  in the quark-antiquark sector. We then study
 spin-spin and spin-orbit mixing, resulting in e.g. quadrupole moments of vector mesons.
For the light mesons, we show how to  extend the famed t$^\prime$Hooft interaction to the light front,  which solves the U(1)
problem and helps produce a light pion. 
 We use the ensuing  light front wavefunctions, to derive the pertinent parton distribution functions, parton amplitudes
and low energy constants. 
\end{abstract}

\maketitle

\section{Introduction}
 The physics of hadrons is firmly based in  Quantum Chromodynamics,  a theory over half a century old.
 One might think that by now this subject has reached a solid degree of maturity with most issues settled.
 Yet persisting tension remains between the non-perturbative aspects of the theory and empirical measurements
 using inclusive and exclusive processes.
 
More specifically, first principle approaches -- lattice and semi-classics -- are focused on the ground state properties 
of the QCD vacuum,  using Euclidean  time formulation. 
 Hadrons are then studied via certain correlation functions (a brief review will be given in  the next subsection).
 However,  a significant part of the experimental information -- parton distribution functions (PDFs) used in deep inelastic inclusive processes, 
 and distribution amplitudes (DAs) used for exclusive processes --
 are defined in the light front kinematics, and therefore are not directly accessible by the Euclidean
 formulation.  Only recently, the first attempt to formulate the appropriate kinematical limits~\cite{Ji:2013dva},   and use
 the lattice for calculating the PDFs~\cite{Zhang:2017bzy,Alexandrou:2018pbm} were carried out with some success.
 
 Bringing the two sides of hadronic physics together is not just a technical issue related with kinematics.
 Even the main pillars of the theory -- confinement and chiral symmetry breaking -- become contentious. In particular,
 60 years ago Nambu and Jona-Lasinio (NJL) \cite{Nambu:1961tp} have  explained that
 pions are light because they are near-massless vacuum waves due to the spontaneous breaking of chiral symmetry.
 The mechanism creating the vacuum quark condensate and the ensuing organization using chiral perturbation theory,  have since
 been discussed and confirmed in countless papers. 
 
 More importantly, the QCD vacuum  in the mesoscopic 
 limit, reveals a multitude of  multi-quark correlations captured by  universal spectral fluctuations in the {\em Zero Mode Zone} (ZMZ)~\cite{Verbaarschot:1993pm}.
  They are analogous to the universal conductance fluctuations  around Fermi surfaces in dirty metals~\cite{Montambaux:1997svv}. We regard 
  these mesoscopic fluctuations as  strong  evidence, in support of the topological
 origin of the spontaneous breaking of chiral symmetry in QCD. Most of the current hadronic models fail to reproduce these fluctuations.
 
 And yet, parton dynamics is still treated as if the vacuum is ``empty" and 
 quarks are treated as massless. There are even suggestions that on the light front,
 there are no condensates~\cite{Brodsky:2009zd,Brodsky:2012ku}. 
 The  pion was also suggested to be massless due to other reasons~\cite{DeTeramond:2021jnn}.
Recently these arguments were revisited~\cite{Ji:2020baz}, and ``quasi-PDF" 
have been  calculated on the lattice~\cite{Ji:2020ect} (and references therein), obviously with all nonperturbative effects included.
 
 Still, there remains a significant gap between light-front observables used and hadronic spectroscopy
 (as well as atomic and nuclear ones): the former focuses  on  certain {\em matrix elements}
 (DAs, PDFs, TMDs, etc), rather than the wave functions,  or the underlying Hamiltonian. This approach is entirely driven by
 information deduced from experiment. 
 
 Indeed, one can calculate various inclusive and exclusive reactions using DAs. But
 their number  is in principle infinite, as there  are 
unlimited number of operators. (The standard approach  is to include only those with the smallest values of their $twist$, 
dimension minus spin, thereby  limiting the discussion to the large $Q^2$ domain.)  The normalization
of the DAs is done via a number of empirical constants like $f_\pi,f_\rho$. 

Hadronic spectroscopy, as  in many other similar fields, goes in the opposite direction, from
the underlying theory to effective Hamiltonians, to wave functions, to matrix elements. 
Indeed, one Hamiltonian produces many eigenstates, with well defined
 wave functions, naturally normalized and mutually orthogonal. 
 From them any  number of matrix elements of interest can be calculated.  

The light-front wave functions were classified in well-known papers such as \cite{Ji:2002xn}, but
hardly used. Only for the pion -- a very special particle, a Nambu-Goldstone mode -- 
there is determination  of both its components, from model-dependent Bethe-Salpeter 
equations~\cite{Shi:2015esa,Shi:2018zqd}, and  from quasi-DAs in the instanton vacuum~\cite{Kock:2021spt}.

Model Hamiltonians were invented,
but not related to the underlying physics. 
The spin-dependent forces -- so important in spectroscopy --
have not been included.
In~\cite{Shuryak:2021fsu} we reviewed  their perturbative and  non-perturbative 
aspects in the rest frame, and 
in~\cite{Shuryak:2021hng} we showed how to extend 
the non-perturbative contributions to the light front.

In this paper, a comprehensive derivation of the perturbative spin contributions
will be given using Wilson lines on the light front. When combined with the non-perturbative
contributions from~\cite{Shuryak:2021hng}, it provides a first principle Hamiltonian on the 
light front. The spectroscopic implications of this Hamiltonian for heavy and light mesons will 
be investigated. 
 
The organization of the paper is as follows: in section section~\ref{sec_II}-\ref{sec_III} we give a first principle derivation of the light front Hamiltonian,
through an analytical continuation of pertinent Wilson loops from Euclidean to Minkowski signature. The derivation
includes both the perturbative and non-perturbative gluonic contributions in the QCD vacuum at low resolution. In section~\ref{sec_IV}
we limit the light front Hamiltonian to the contributions stemming from confinement and Coulomb, and analyze their role on heavy quarkonia,
with Upsilonium as an example. In section~\ref{sec_V} we briefly review how parity is  defined  on the light front, and how it is used to
organize the light front wave-functions for mesons. In section~\ref{sec_VI} we consider the mixing induced by the tensor contribution
to the light front Hamiltonian, onto heavy quarkonia. We show that the quadrupole moment of Upsilonium on the light front is about comparable
to the one extracted from other approaches both at rest and also on the light front. In section~\ref{sec_VII}, we consider the additional mizing
induced by spin-orbit coupling on the light front, and apply in this case to the light meson spectrum. In section~\ref{sec_THH} we show how
the the subtle zero-modes associated to tunneling through instantons in Euclidean signature are lifted to the light front, using the LSZ
reduction in coordinate space. We use it to derive the famed t$^\prime$Hooft interaction on the light front. 
In section~\ref{sec_OBSERVABLES} we use
our light front wavefunctions  to derive the parton distributions functions and amplitudes  of heavy and light mesons, and their pertinent low energy constants.
The  extraction of the mesonic form factors, is also briefly discussed.
 Our conclusions are in section~\ref{sec_CON}. A number of Appendices are added to complement some of the results in the text.

\begin{figure}[h!]
	\begin{center}
		\includegraphics[width=8cm]{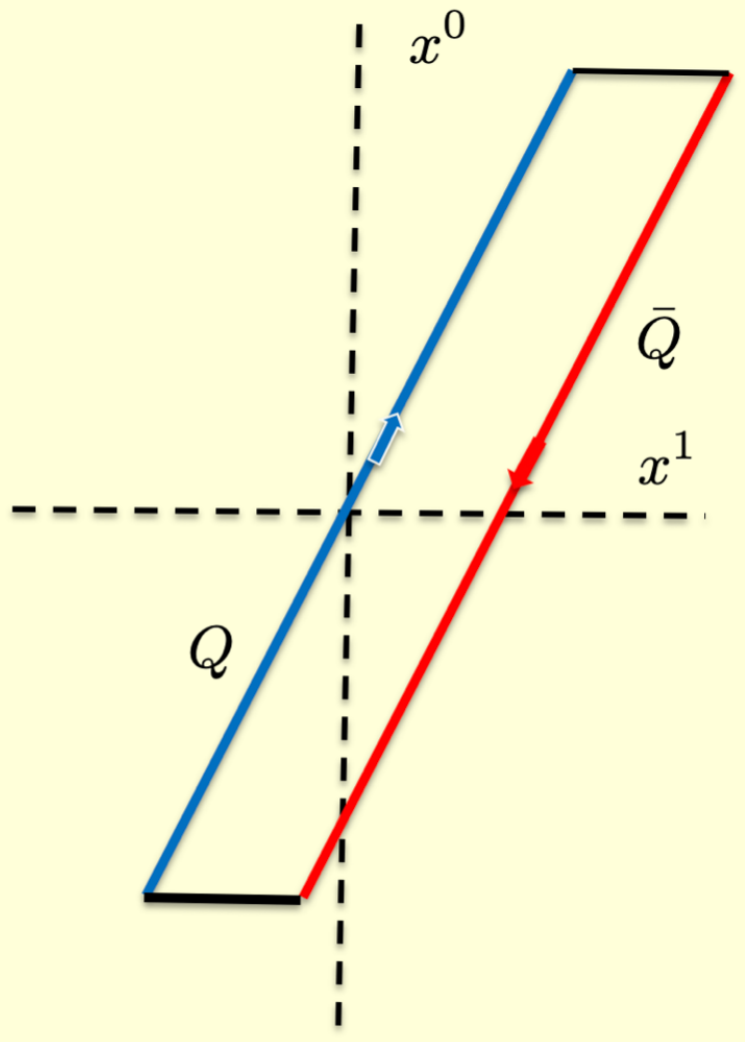}
		\caption{Wilson loop for a $\bar QQ$ meson on the light-front.}
		\label{fig_wilson_meson}
	\end{center}
\end{figure}

\section{Perturbative light-front Hamiltonian, via analytic continuation from Euclidean amplitudes}~\label{sec_II}

 In the infinite momentum frame, a meson state composed of a quark and antiquark $Q\bar Q \equiv Q_1Q_2$  is characterized by the closed Wilson loop
or a {\it dipole}, sloped  along the light cone with rapidity $\chi$ as shown in Fig.~\ref{fig_wilson_meson}.    The same Wilson loop follows from the 
Euclidean Wilson loop  at an angle $\theta$ by analytical continuation $\theta\rightarrow -i\chi$, as we discussed 
in the second paper of this series~\cite{Shuryak:2021hng}. This construction follows
the  original suggestion for quark-quark scattering in~\cite{Meggiolaro:1997mw}, and  its extension to {\it dipole-dipole} scattering
in the QCD vacuum~\cite{Shuryak:2000df,Giordano:2009su}, many years ago. The same construction was used in the holographic context, to address 
hadron-hadron  scattering in the Regge limit~\cite{Janik:2000aj,Basar:2012jb,Shuryak:2017phz}.

With this in mind, the result is the squared meson mass operator, or light front Hamiltonian $H_{LF}$
 
\ba
\label{16Z} 
H_{LF}\approx && \frac{k_\perp^2+m_Q^2}{{x\bar x}} + 2P^+P^-\nonumber\\
\approx &&
 \frac{k_\perp^2+m_Q^2}{{x\bar x}} 
+ 2M (\mathbb V_{Cg}(\xi_x)+\mathbb V_C(\xi_x)\nonumber\\
&&+\mathbb V_{SD}(\xi_x, b_\perp)
 +\mathbb V_{TH}(\xi_x, b_\perp)) 
\ea
The non-perturbative contributions in~(\ref{16Z}) were discussed in~\cite{Shuryak:2021hng},
along with the ordering ambiguities.
The perturbative contributions will be derived below.
On the light front, the  invariant distance $\xi_x$ is

\bea
\xi_x=\bigg(\bigg|\frac{id/dx}{M}\bigg|^2+{b_\perp}^2\bigg)^{\frac 12}
\eea
with  longitudinal distance $\gamma b_3=id/dx/M$, the  conjugate of Bjorken-x or
$x=k^3/P^3$.
The explicit $\gamma$-factor compensates the Lorentz contraction along the 3-direction.

\subsection{Wilson lines dressed by spin variables}

The perturbative contribution to the central potential on the light front,  induced by a  one-gluon exchange with an effective mass $m_G$ in the RIV, can be constructed  using the general technique of a sloped Wilson loop as we detailed in~\cite{Shuryak:2021hng}. In particular, the
one-gluon  interaction with spin effects,  follows by dressing the Wilson loop or holonomies in Fig.~\ref{fig_wilson_meson}, with explicit spin factors

\begin{widetext}
\bea
\label{ONEX}
\bigg<{\rm Tr}{\bf P}&&\bigg[{\rm exp}\bigg(+g\int d\tau_1(i\dot{x}(\tau_1)\cdot A(x(\tau_1)+\frac 1{4}\sigma_{1\mu\nu}F_{\mu\nu}(x(\tau_1))\bigg)
\nonumber\\
&&\times {\rm exp}\bigg(-g\int d\tau_2(i\dot{x}(\tau_2)\cdot A(x(\tau_2)+\frac 1{4}\sigma_{2\mu\nu}F_{\mu\nu}(x(\tau_2))\bigg)\bigg]\bigg>
\eea
\end{widetext}
with $\sigma_{\mu\nu}=\frac 1{2i}[\gamma_\mu, \gamma_\nu]$,  and $\sigma_{\mu\nu}=\eta_{a\mu\nu} \sigma^a$ using $^\prime$t Hooft symbol. 
The averaging is understood using the QCD action.
We have made explicit the gauge coupling $g$, for a perturbative treatment to follow.

For  massive quarks traveling on straight trajectories, the affine time
$\tau$ relates to the conventional time  $t$ through 

\bea
\label{SS2}
\mu=\frac {dt}{d\tau}=\frac{m_Q}{\sqrt{1+{\dot{\vec x}}^2}}\rightarrow \gamma m_Q
\eea
in Euclidean signature. We note that the holonomies tracing out the Wilson loop in Fig.~\ref{fig_wilson_meson} are  unaffected by the exchange $\tau\rightarrow t$,
in contrast to the spin contributions which get rescaled by $1/\mu$. This will be exploited below.

\subsection{One-gluon exchange and the Coulomb interaction}

The Coulomb interaction between a $Q\bar Q\equiv Q_1Q_2$ pair attached to the Wilson lines, can be obtained
in perturbation theory by expanding the holonomies, and averaging the $AA$ correlator in leading order. 
For that, we parametrize the world-lines by

\bea
x_\mu(t_1)&=&(0,0, {\rm sin}\theta\,t_1, {\rm cos}\theta\, t_1)\qquad\nonumber\\
x_\mu(t_2)&=&(b_1,b_2, {\rm sin}\theta\, t_2+b_3, {\rm cos}\theta\, t_2)
\eea
The perturbative one-gluon contribution 
from (\ref{ONEX}) reads

\begin{widetext}
\bea
\label{ONE1}
g^2 T_1^AT_2^B\int\,dt_1\int\,dt_2
\bigg({\rm cos^2}\theta \langle A_4^A(t_1)A^B_4(t_2)\rangle+{\rm sin^2}\theta \langle A_3^A(t_1)A^B_3(t_2)\rangle
+2{\rm sin}\theta {\rm cos}\theta  \langle A_4^A(t_1)A^B_3(t_2)\rangle\bigg)\nonumber\\
\eea
\end{widetext}
with the gluon correlator in Feynman gauge

\bea
\label{ONE2}
\langle A_\mu^A(t_1)A^B_\nu(t_2)\rangle=\frac 1{2\pi^2}\frac{\delta^{AB}\delta_{\mu\nu}}{|x(t_1)-x(t_2)|^2}
\eea
Inserting (\ref{ONE2}) into (\ref{ONE1}) and changing variables $T_E=t_1+t_2$ and $\tau=t_1-t_2$, yield

\bea
\label{ONE3}
&&\frac{g^2 T^A_1T_2^A}{2\pi^2}\int\,\frac {dT_E}2\int\,dt
\frac 1{t^2+{\rm cos}^2\theta\,b_3^2+b_\perp^2}\nonumber\\
&&=\frac{g^2T_1^AT_2^A}{4\pi} \frac {T_E}{\sqrt{b_3^2{\rm cos}^2\,\theta+b_\perp^2}}
\eea
The analytical continuation $\theta\rightarrow -i\chi$ and $T_E\rightarrow iT_M$ of (\ref{ONE3}), re-exponentiates to the Coulomb contribution

\bea
{\rm exp}\bigg[-i\gamma T_M\bigg(-\frac{g^2T_1^AT_2^A}{4\pi} \frac {1/\gamma}{\sqrt{\gamma^2b_3^2+b_\perp^2}}\bigg)\bigg]
\eea
with $\gamma T_M$ the dilatated time along the light-like Wilson loop. The Coulomb contribution to the light front $Q\bar Q$ Hamiltonian $P_{Cg}^-$ follows, leading the squared invariant mass as

\bea
\label{ONE4}
&&2P^+P_{Cg}^-=
2P^+\bigg(-\frac{g^2T_1^AT_2^A}{4\pi} \frac {1/\gamma}{\sqrt{\gamma^2b_3^2+b_\perp^2}}\bigg)\nonumber\\
&&\rightarrow 2M\bigg(-\frac{g^2T_1^AT_2^A}{4\pi} \frac{1}{\xi_x}\bigg)=2M\mathbb V_{Cg}(\xi_x)
\eea
with $P^+/M=\gamma$, and $\gamma b_3\rightarrow id/dx/M$ the conjugate of Bjorken-x.

 In the random instanton vacuum (RIV), the perturbative gluons acquire a momentum dependent mass
 from their rescattering through the instanton-anti-instanton ensemble~\cite{Musakhanov:2021gof} 

\bea
m_G(k\rho)&=&m_G\, \bigg(k\rho\,K_1(k\rho)\bigg)\nonumber\\
 m_G\rho &\approx& 2\bigg(\frac {6\kappa }{N_c^2-1}\bigg)^{\frac 12}\approx 0.55
\eea
using the estimate $\kappa=\pi^2\rho^4 n_{I+\bar I}$ in the right-most result. With this in mind, (\ref{ONE4}) is now

\begin{widetext}
\bea
\label{ONEG}
\mathbb V_{Cg}(\xi_x)=-\frac{g^2 T_1^AT_2^A}{2\pi^2}  \frac 1{\xi_x}\int_0^\infty\frac {dx\,x{\rm sin}x}{x^2+(\xi_xm_G(x\rho/\xi_x))^2}         
\rightarrow -\frac{g^2 T_1^AT_2^A}{4\pi}\frac {e^{-m_G \xi_x}}{\xi_x}
\eea
\end{widetext}
with the right-most result following for a constant gluon mass.

\subsection{Spin-spin interaction}

The perturbative spin-spin interaction follows from  the cross term  in (\ref{ONEX})

\bea
\label{SS1}
-\frac {g^2}{16}\int d\tau_1 d\tau_2 \langle \sigma_{1\mu\nu}F_{\mu\nu}(x(\tau_1))\sigma_{2\alpha\beta}F_{\alpha\beta}(x(\tau_2))\rangle\nonumber\\
\eea
Note that the  perturbative electric field is purely imaginary in Euclidean signature, leading mostly to phases and not potentials in the long time
limit.  Also,  the Dirac representation $\sigma_{4i}$ is off-diagonal, an indication that the electric contribution mixes
particles and anti-particles, which is excluded by the use of straight Wilson lines on the light front.  With this in mind and using  (\ref{SS2}),  we can reduce (\ref{SS1})  to

\bea
\label{SS3}
&&-\frac{g^2}{4\mu^2}\int dt_1 dt_2 \langle \sigma_{1ij}F_{ij}(x(t_1))\sigma_{2kl}F_{kl}(x(t_2))\nonumber\\
&&=-\frac{g^2}{4\mu^2}\sigma_1^a\sigma_2^b\int dt_1 dt_2 \langle B_a(x(t_1))B_b(x(t_2))\rangle\nonumber\\
\eea
with

\bea
\label{SS4}
&&\big<B_a(x(t_1))B_b(x(t_2))\big> \nonumber\\
&&=T_1^AT_2^B\epsilon_{aij}\epsilon_{bkl}\partial_{1i}\partial_{2k}
\big<A^A_j(x(t_1)A_m^B(x(t_2))\big>\nonumber\\
&&=T^A_1T^A_2(\delta_{ab}\delta_{ik}-\delta_{ak}\delta_{bi})\partial_{1i}\partial_{2k}\nonumber\\
&&\,\,\,\,\,\,\times\frac 1{2\pi^2}\frac 1{|x(t_1)-x(t_2)|^2} 
\eea
Inserting (\ref{SS4}) into (\ref{SS3}), and carrying the time integrations along the sloped Wilson loop
in Fig.~\ref{fig_wilson_meson} give among others

\bea
\label{SS5}
&&-\frac{g^2T_1^AT_2^A}{4\pi} \frac {T_E}{4\mu^2}\nonumber\\
&&\times
\bigg[\sigma_{1\perp}\cdot \sigma_{2\perp}\,\bigg(-3 {\rm cos}^2\theta \frac{({\rm cos}\theta \,b_3)^2}{\xi_\theta^2}+{\rm cos}^2\theta\bigg)\frac 1{\xi_\theta^3}\bigg]\nonumber\\
\eea
which is the dominant contribution under the analytical continuation $\theta\rightarrow -i\chi$, $\mu\rightarrow \gamma m_Q$ and
$T_E\rightarrow iT_M$, in the ultra-relativistic limit  $\gamma\gg 1$, and  in Minkowski signature. The final spin-spin contribution to the squared mass is in
general

\bea \label{eqn_VSS_pert}
H_{SS}&=&2M\bigg[\frac {\sigma_{1\perp}\cdot\sigma_{2\perp}}{4m_{Q1}m_{Q2}}
\bigg( \nabla_\perp^2 \mathbb V_{Cg}(\xi_x)\bigg)\bigg]\nonumber\\
&=&2M\mathbb V_{SS}(\xi_x, b_\perp)\nonumber\\
\eea

\subsection{Spin-orbit interaction}

\subsubsection{Cross spin-orbit}

The  cross spin-orbit interaction is readily obtained   from the $12+21$ cross terms

\bea
&&-\frac{g^2 T_1^AT_2^B}{2}\,\sigma_2^a\,\nonumber\\
&&\times\int d\tau_1 d\tau_2 \, i\dot{x}_i(\tau_1)\big<A_i^A(x(\tau_1)B_a^B(x(\tau_2))\big>\nonumber\\
&&+ 1\leftrightarrow  2
\eea
which can be reduced to

\bea
&&-\frac{ig^2T_1^AT_2^B{\rm sin}\theta}{2\mu}\sigma_2^a s_1\nonumber\\
&&\times\int dt_1 dt_2\langle A^A_3(x(t_1))B^B_a(x(t_2))\rangle\nonumber\\
&&+1\leftrightarrow 2
\eea
with  $s_{1,2}={\rm sgn}(v^3_{1,2})$  the signum of the 3-velocity of particle 1,2
(a more refined definition will be given below). 
After carrying the integrations, and the analytical continuations, the spin-orbit contribution to the squared mass is

\begin{widetext}
\bea
\label{MLS12}
H_{SL,12}=2M\bigg[\bigg(\frac{\sigma_2\cdot (b_{12}\times s_1\hat 3)}{2m_{Q2}}
-\frac{\sigma_1\cdot (b_{21}\times s_2\hat 3)}{2m_{Q1}}\bigg)
\bigg(\frac 1{\xi_x}\mathbb V^\prime_{Cg}(\xi_x)\bigg)\bigg]
\eea
\end{widetext}
in general, with $b_{21}=-b_{12}\equiv b_\perp$.

\subsubsection{Standard spin-orbit}

The standard self spin-orbit interaction with Thomas precession is more subtle.  To unravel it,  we note that 
the insertion of a single spin contribution along the path-ordered Wilson loop amounts to expanding
the spin factors in (\ref{ONEX}) to first order,  and retaining the holonomies to all orders in ${\bf 1}_\theta$,
namely

\bea
\label{LS1}
 \frac 1{4\mu} \sigma_{1\mu\nu}\int dt_1 \langle gF_{\mu\nu}(x(t_1))\,{\bf 1}_\theta\rangle + 1\leftrightarrow 2\nonumber\\
\eea
with the path ordered color-spin  trace subsumed. Here ${\bf 1}_\theta$ refers to the slated Wilson loop in Fig.~\ref{fig_wilson_meson} without the spin dressing.
We now decompose

\bea
\label{LS2}
F_{\mu\nu}=v_{1\mu} v_{1\alpha} F_{\alpha\nu} +F_{\mu\nu}^\perp\equiv F_{\mu\nu}^{||}+F_{\mu\nu}^\perp
\eea
into a contribution parallel to $v_{1}=\dot{x}_1$ and a contribution orthogonal to $v_1$. The contribution 
parallel to the worldline when inserted in (\ref{LS1}) can be undone by the identity
(see Eq.71 in~\cite{Shuryak:2021hng})

\bea
&&\int dt_1 \langle gv_\alpha F_{\alpha\nu}(x(t_1))\,{\bf 1}_\theta\rangle\nonumber\\
&& =-i\partial_{1\nu}\langle {\bf 1}_\theta\rangle
\equiv -i\partial_{1\nu}\,e^{-T_E\mathbb V_C({\xi_\theta})}
\eea
with $\mathbb V_C(\xi_\theta)\approx \mathbb V_{Cg}(\xi_\theta)$ the central Coulomb potential in perturbation theory. The 
longitudinal contribution to (\ref{LS1}) 

\bea
\label{LS3}
- \frac 1{4\mu} \sigma_{1\mu\nu}v_{1\mu} i\partial_{1\nu}\,e^{-T_E\mathbb V_C({\xi_\theta})} + 1\leftrightarrow 2\nonumber\\
 \eea
is gauge-invariant. After carrying the analytical continuation, (\ref{LS3}) contributes both a real and imaginary part. The latter 
is an irrelevant  phase factor in Euclidean signature. The  {\it real } part contributes to the direct mass squared operator as

\begin{widetext}
\bea
\label{MLS11}
H_{LS, 11}=
2M\bigg[\bigg(\frac{\sigma_1\cdot (b_{12}\times s_1\hat 3)}{4m_{Q1}}
-\frac{\sigma_2\cdot (b_{21}\times s_2\hat 3)}{4m_{Q2}}\bigg)
\bigg(\frac 1{\xi_x}\mathbb V^\prime_{Cg}(\xi_x)\bigg)\bigg]
\eea \label{eqn_pert_spin_orb}
\end{widetext}
in leading order in perturbation theory.
This is the standard spin-orbit contribution with the correct Thomson correction on the light front, familiar from atomic physics in the rest frame.
The total perturbative spin  contribution on the light front, is the sum of  (\ref{eqn_VSS_pert}), (\ref{MLS12}) and (\ref{MLS11}),

\begin{widetext}
\bea
\label{PERP1}
H_{LS,g}&=&2M\bigg(
\frac{l_{1\perp}\cdot S_{1\perp}}{2m_{Q1}^2}-\frac{l_{2\perp}\cdot S_{2\perp}}{2m_{Q2}^2}
+\frac{l_{1\perp}\cdot S_{2\perp}}{m_{Q1}m_{Q2}} -\frac{l_{2\perp}\cdot S_{1\perp}}{m_{Q1}m_{Q2}}\bigg)
\frac 1{\xi_x}\mathbb V_{Cg}^\prime(\xi_x)\nonumber\\
&+&2M\bigg(\frac {S_{1\perp}\cdot S_{2\perp}}{m_{Q1}m_{Q2}}\bigg)
 \nabla_\perp^2 \mathbb V_{Cg}(\xi_x)
\eea
with the respective spins $\vec S_{1,2}=\vec \sigma_{1,2}/2$, and transverse  orbital momenta

\bea
\label{PERP2}
l_{1,2\perp}=\pm (b_\perp\times m_{Q1,2}s_{1,2}\hat 3)_\perp\qquad 
 s_{1,2}={\rm sgn}(v_{1,2}) \rightarrow \frac{Mx_{1,2}}{m_{Q1,2}}
 \eea

\section{Instanton  contributions to Wilson line amplitudes}~\label{sec_III}

\subsection{Central potential}

The central potential-operator induced by instantons is given by

\be
\label{CENT1}
\mathbb V_C(\xi_x)= \bigg(\frac{4\kappa }{N_c \rho}\bigg){\bf H}(\tilde\xi_x)
\ee
with  the integral operator 

\bea
\label{W10X}
{\bf H}(\xi_{x} )=&&\int_0^\infty y^2 dy\int_{-1}^{+1} dt \nonumber\\
&&\times\bigg[1-{\rm cos}\bigg(\frac{\pi y}{\sqrt{y^2+1}}\bigg)
{\rm cos}\bigg(\pi\bigg(\frac{y^2+\tilde\xi_x^2+2\xi_{x}yt}{y^2+\tilde\xi_x^2+2\tilde\xi_{x}yt+1}\bigg)^{\frac 12}\bigg)\nonumber\\
&&-\frac{y+\xi_{x}t}
{(y^2+\xi_x^2+2\tilde\xi_{x}yt)^{\frac 12}}
{\rm sin}\bigg(\frac{\pi y}{\sqrt{y^2+1}}\bigg)
{\rm sin}\bigg(\pi\bigg(\frac{y^2+\tilde\xi_x^2+2\tilde\xi_{x}yt}{y^2+\xi_x^2+2\tilde\xi_{x}yt+1}\bigg)^{\frac 12}\bigg)\bigg]
\eea
\end{widetext}
with the dimensionless invariant distance on the light front $\tilde\xi_x=\xi_x/\rho$. ${\bf H}(\tilde\xi_x)$ admits the short  distance limit

\bea
\label{17Z} 
&&{\bf H}(\tilde\xi_x) \approx + \bigg(\frac{\pi^3}{48}-\frac{\pi^3}3J_1(2\pi )\bigg)\tilde\xi_x^2\nonumber\\
&&+
\bigg(-\frac{\pi^3(438+7\pi^2)}{30720}+\frac{J_2(2\pi)}{80}\bigg)\tilde\xi_x^4
\eea
and large distance limit
\bea
{\bf H}(\xi_x) \approx -\frac{2\pi^2}3\bigg(\pi J_0(\pi)+J_1(\pi)\bigg)+\frac{C}{\tilde\xi^p_x}
\eea
with $p\ll 1$ and $C>0$. The  large asymptotic is to be subtracted in the definition of the potential. This will be subsumed throughout.
In the dense instanton vacuum discussed in~\cite{Shuryak:2021hng}, the central potential (\ref{CENT1}) is almost linear 
at intermediate distances $0.2-0.5$ fm. At larger distances,  the linearly confining potential with string tension $\sigma_T$ takes over,
in good agreement with most  lattice simulations.

\subsection{Spin dependent potentials}

On the light front, the spin-dependent interactions  captured by $V_{SD}$ and due to the non-zero modes
in (\ref{16Z}),  have been discussed in general in~\cite{Shuryak:2021hng}, with the results

\begin{widetext}
\bea
\label{SPINFULL}
\mathbb  V_{SD}(\xi_x, b_\perp)=&&+
\bigg[ \frac{\sigma_1\cdot ( {b}_{12}\times s_1\hat 3)}{4m_{Q1}}
-\frac{\sigma_2\cdot ( {b}_{21}\times s_2\hat 3)}{4m_{Q2}} \bigg]\,\frac 1{\xi_x}\mathbb V_C^\prime(\xi_x)\nonumber\\
 &&+
\bigg[\frac{\sigma_1\cdot (b_{12}\times s_1\hat 3)}{2 m_{Q1}}-\frac{\sigma_2\cdot (b_{21}\times s_2\hat 3)}{2 m_{Q2}}\bigg]
\frac 1{\xi_x}\mathbb V_1^\prime(\xi_x)\nonumber\\
&&+
\bigg[\frac{\sigma_2\cdot(b_{12}\times s_1\hat 3)}{2 m_{Q2}}-\frac{\sigma_1\cdot(b_{21}\times s_2\hat 3)}{2 m_{Q1}}\bigg]\frac 1{\xi_x}
\mathbb V_2^\prime(\xi_x)\nonumber\\
&&+
\bigg[\frac 1{4m_{Q1}m_{Q2}}\bigg[\bigg(\sigma_{1\perp}\cdot\hat b_{21}\sigma_{2\perp}\cdot\hat b_{21}-\frac 12\sigma_{1\perp}\cdot\sigma_{2\perp}\bigg)\bigg]\mathbb V_3(\xi_x)
\eea
\end{widetext}
with again $b_{21}=-b_{12}\equiv b_\perp$.
All potentials follow from the central  instanton potential $\mathbb V_C(\xi_x)$, thanks to self-duality 

\bea
\mathbb V_1(\xi_x)&=&\mathbb V_2(\xi_x)-\mathbb V_C(\xi_x)=-\frac 12 \mathbb V_C(\xi_x)\nonumber\\
\mathbb V_2(\xi_x)&=&+\frac 12 V_C(\xi_x)\nonumber\\
\mathbb V_3(\xi_x)&=&+2 \frac{b_\perp^2}{\xi_x^2}\mathbb V_C^{\prime\prime}(\xi_x)
\eea
As a result, the first and second contribution in (\ref{SPINFULL}) cancel out,
leaving only the cross spin orbit plus tensor contributions in the instanton vacuum,

\begin{widetext}
\bea \label{eqn_inst_spin_orb}
H_{LS}=2M\bigg(
\bigg(\frac{l_{1\perp}\cdot S_{2\perp}}{m_{Q1}m_{Q2}} -\frac{l_{2\perp}\cdot S_{1\perp}}{m_{Q1}m_{Q2}}\bigg)
\frac 1{\xi_x}\mathbb V_C^\prime(\xi_x)
+\frac 1{m_{Q1}m_{Q2}}(S_{1 \perp}\cdot \hat{b}_\perp S_{2\perp} \hat{b}_\perp-\frac 12 S_{1\perp}\cdot S_{2\perp})
\frac {2b_\perp^2}{\xi_x}\mathbb V_C^{\prime\prime}(\xi_x)\bigg)\nonumber\\
\eea
\end{widetext}
with $\vec l_{1,2}$ given in (\ref{PERP2}), and $V_C(\xi_x)$ in (\ref{CENT1}).
The contributions stemming from the zero modes are not included. They will be discussed separatly below.


\section{Heavy quarkonia on the light front}~\label{sec_IV}

In our second paper~\cite{Shuryak:2021hng} we introduced ``the basic problem" of meson structure, of two
constituent quarks connected by a classical relativistic string, which was then studied using
both a semiclassical approach,  and  a relativistic Klein-Gordon equation. Our main focus there
was on the correspondence between the conventional  
treatment in the rest frame,  versus  the analysis on the light front using the Hamiltonian we derived.
Of course, frame-invariant quantities --masses in particular -- obtained in both ways must agree. We specifically investigated 
the  linear rise of the Regge trajectories, with the principal quantum number $M_n^2\sim n$
 (not angular momentum). 
 
 In this paper we will carry out a larger set of studies, including
 not only the confining string, but also  various other terms in the Hamitonian.
 In particular, the perturbative (Coulomb)
 term, and  most importantly, the terms containing spin and orbital momentum variables.
 In doing so, it is also natural to widen the set of applications. Therefore 
 here we start with heavy quarkonia, before returning to the  light quark systems.
 
 In the quarkonia settings,  we  can use the
 large quark mass  as an extra parameter, to discriminate between distinct physical contributions.
 Remarkably, on the light front all meson problems,
 from  bottomonia to pions, can be studied in essentially the same setting, just
 by changing the mass value.

\subsection{Excited states of bottomonium via the Schroedinger equation in the rest frame}

Let us start  by  focusing  first on  heavy quarkonia. Such an
approach is more convenient in this work, devoted to the
mixing between states with different spin and orbital momenta.
In  heavy quarkonia these relativistic effects are naturally suppressed by the
nonrelativistic motion of heavy quarks (or, in other terms, their small magnetic moments
$\mu\sim g/m_Q$). 

\begin{figure}[h!]
\begin{center}
\includegraphics[width=7cm]{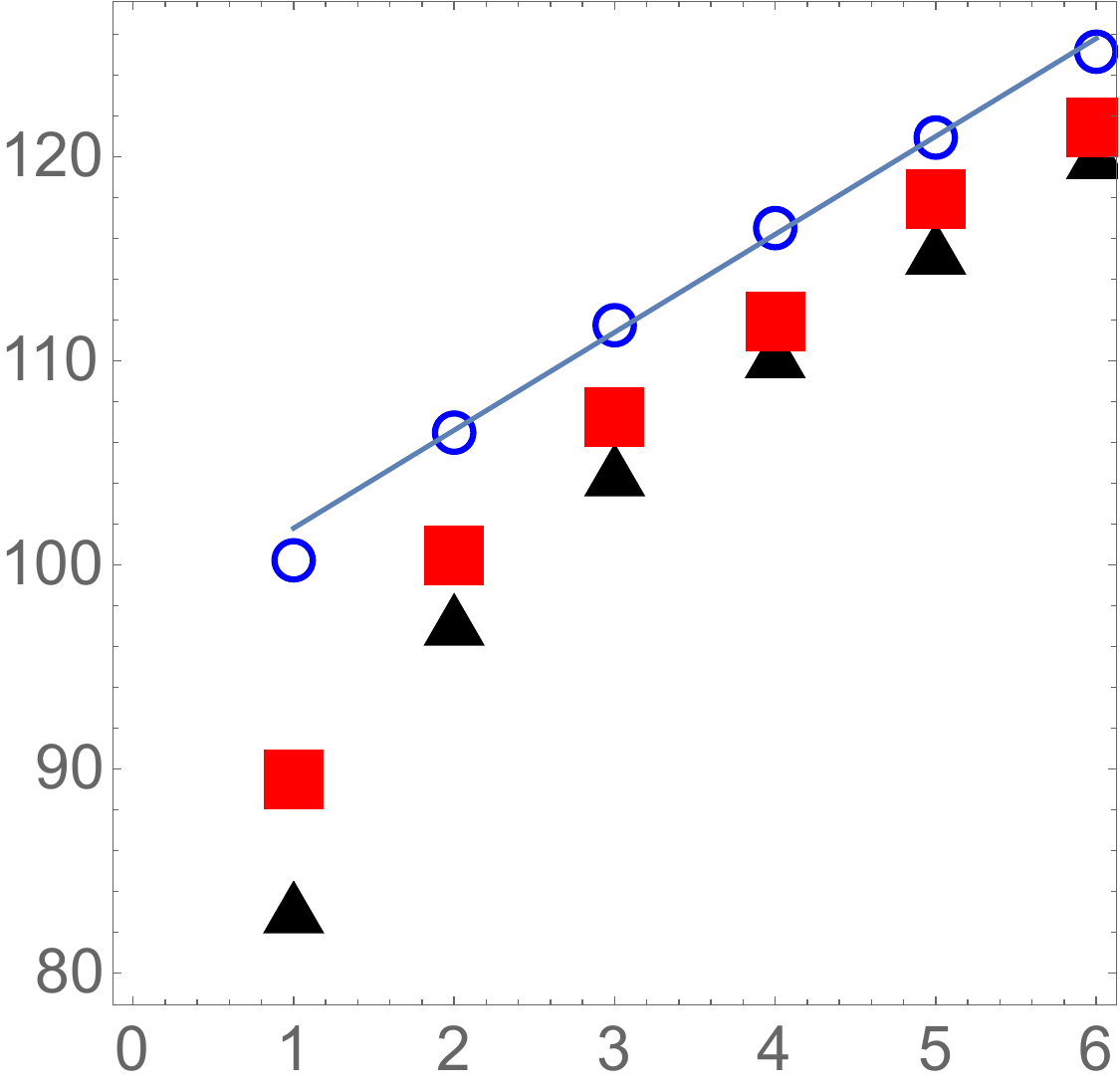}
\caption{$M_{n+1}^2\, (GeV)^2$ versus $n+1,n=0,..5$, for the six $S$ zero orbital momentum (L=0) states of  bottomonium.
The red squares correspond to the experimentally observed Upsilons. The black triangles 
show the masses obtained from the Schroedinger equation, with the Cornell potential (linear and Coulomb potentials, no spin forces). 
The blue circles show the masses if 
the Coulomb potential is switched off,  and only  the  linear potential is used. The straight line is shown for comparison. 
 }
\label{fig_6upsilons}
\end{center}
\end{figure}

The first question to be addressed is : how well the heavy quarkonia states can be
represented by linear-linear Regge trajectories?
In Fig.\ref{fig_6upsilons} we show the experimental masses of the $(nS), n=0-5$ Upsilons, 
compared to the standard results from the Schroedinger equation, with the Cornell potential
(black triangles) and with only its linear part $V_{conf}=\sigma_T r$ (blue circles). The first 
observation 
 is that using a linear potential alone (blue circles), we 
 find a nearly-linear Regge trajectory. This observation will be important
 in the next subsection, as it shows that even for heavy bottomonia,  the light-front
 Hamiltonian can be approximated by an oscillator with good accuracy. Note 
however, that the slope of the straight line,  is here completely different 
from the $1/\alpha'$ slope of a similar trajectory for light mesons 
(e.g. for $\omega$ mesons we used in \cite{Shuryak:2021fsu}).

\begin{figure}[h!]
\begin{center}
\includegraphics[width=7cm]{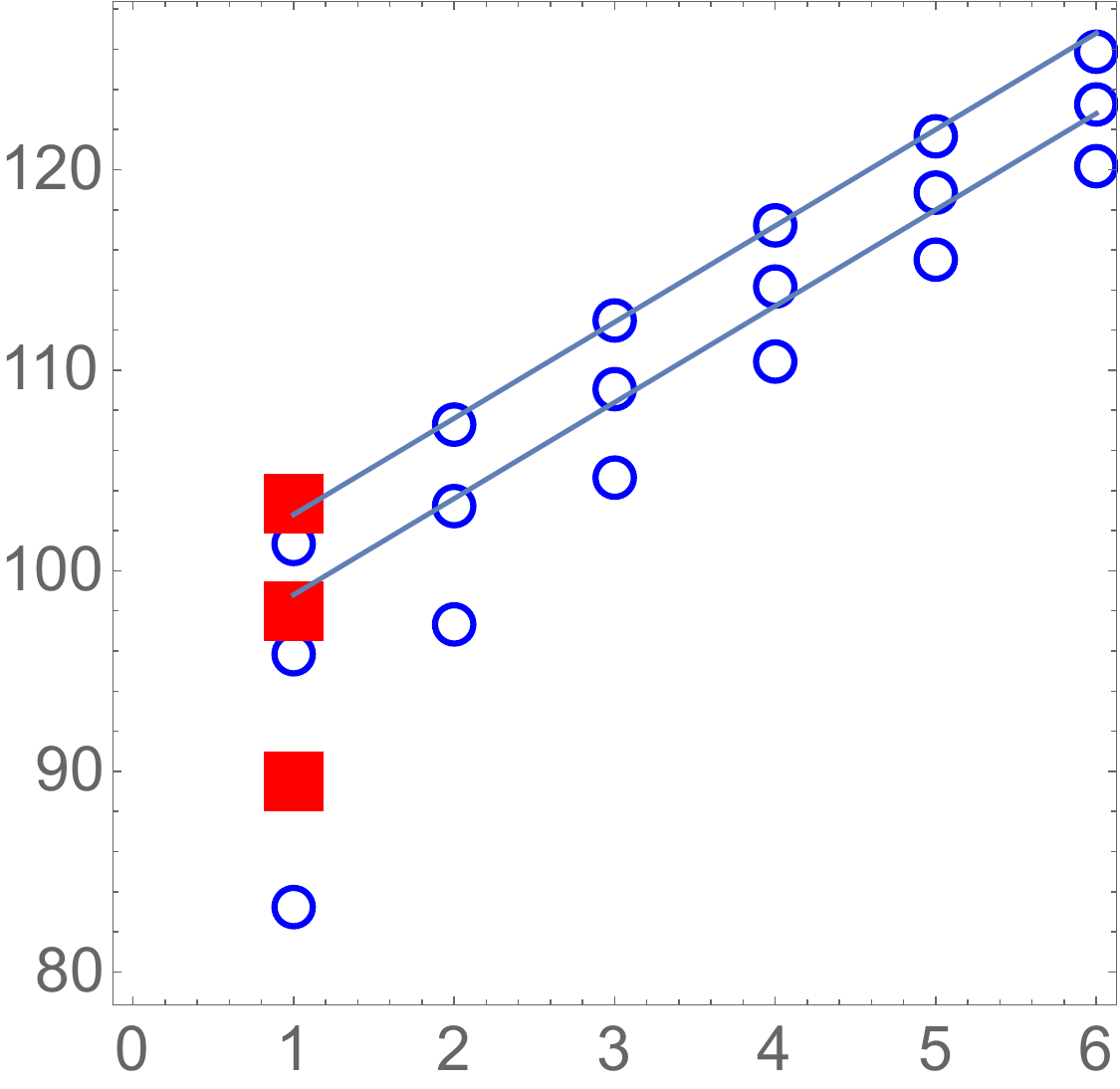}
\caption{$M_{n+1}^2\, (GeV)^2$ versus $n+1, n=0,1...$ for  three families  of  bottomonium
states, with orbital momentum $L=0,1,2$ (from bottom up). 
The red squares correspond to the experimentally observed $\Upsilon,h_b, \Upsilon_2$
mesons (from bottom up). The blue circles show
 masses obtained from the Schroedinger equation,
  with the Cornell potential (linear and Coulomb potentials, still without spin forces).  The straight lines are shown for comparison. 
 }
\label{fig_3L_bb}
\end{center}
\end{figure}

  The second observation is that the expected contribution from the  spin-dependent potential $V_{SS}$ (responsible for
splitting between squares and triangles),  is positive and decreases with $n$. The former is due to the 
positivity of the spin factor $\vec S_1\cdot\vec S_2=1/4$, and the second
to the fact that $V_{SS}(r)$ is rather short range, in comparison to the size
of the lowest Upsilon,  but much smaller than the sizes of  the excited ones.
Another way to anticipate the accuracy of an oscillator approximation in the light front
description (discussed in \cite{Shuryak:2021hng} and using $\omega_3$ mesons with $L=2$),  is to
study the mass dependence of bottomonium on its orbital momentum $L$. 
In Fig.\ref{fig_3L_bb} we show the calculated 18 squared masses for $n=0-5$ 
(left-to-right) and $L=0,1,2$ (bottom-to-top). While the Coulomb potential was included, it affects mostly and only $n=0,L=0$ Upsilon. The Regge trajectories for nonzero angular momentum
$L=1,2$,  show better linear dependence on $n$
 than $L=0$. The  corresponding wave function vanishes at the origin $r=0$, 
 and is less affected by short-range Coulomb and spin-dependent forces.
 
We further note, that  at larger $n$
(right side of the plot) the dependence on $L$ also becomes linear, as the $L=0,1,2$ 
points
become equidistant. This observation  encourages us to think that the oscillator
description of the light front Hamiltonian and LFWFs,  will need only relatively small corrections.  
 
\subsection{Bottomonium on the light front}

In~\cite{Shuryak:2021hng}, we described how we may  include the  linear confining term in $H_{LF}$
(instanton induced at intermediate distances),
and make it  more user friendly, by eliminating the square root
using the well-known einbein $e=1/a$ trick, i.e. 

\bea
\label{EINBEIN}
&&2M \mathbb V_C(a,b,x,b_\perp)\nonumber \\
&&\approx \sigma_T\bigg(\frac{|id/dx|^2+b\, b_\perp^2}a+a\bigg)
\eea
Here $a,b$ are variational parameters.
The minimization with respect to $a$ is assumed, followed by the substitution 
$b\rightarrow M^2\approx (2m_Q)^2$ for heavy mesons,  and most light ones.
(For the pion, this last substitution
is not valid, as we have shown in~\cite{Shuryak:2021hng}). 

For a numerical analysis of (\ref{EINBEIN}), we used in~\cite{Shuryak:2021hng}
a basis set of functions composed of a 2-dimensional transverse oscillator, times
longitudinal states  $sin(\pi n x)$ with odd $n$, as we briefly review in Appendix~\ref{2d_basis}.
More specifically, the light front Hamiltonian can be re-arranged as follows

\be 
\label{HLFX}
H_{LF}=H_0+\tilde V + V_{perp}+V_{spin} 
\ee
with the spin-part including both the perturbative and non-perturbative instanton
contributions. As we noted earlier,  in the dense instanton vacuum, the central part is hardly
differentiable from the linear confining potential at  intermediate distances.

The first contribution  $H_0$ 

\begin{equation}
\label{H0X}
 H_0={\sigma_T \over a} \bigg( -{\partial^2  \over \partial x^2}-b{\partial^2  \over \partial  \vec k_\perp^2} \bigg) + \sigma_T  a + 4(m_Q^2+ k_\perp^2)
\end{equation} 
is diagonal in the functional basis used~\cite{Shuryak:2021hng}. 
In  this form, we make use of  the momentum representation, with $\vec k_\perp$ as variable. Similarly, one can
use the coordinate representation with $\vec b_\perp$ as a variable,  and $ \vec k_\perp=i \partial /\partial \vec b_\perp $.
The latter choice is much more convenient when discussing states with nonzero angular
momenta, in  relation  to the azimuthal angle coordinate $\phi$ in the transverse plane (see more on that in Appendix).

The second contribution $\tilde V$
\begin{equation}  
\tilde V(x,\vec k_\perp)\equiv (m_Q^2+k_\perp^2)\bigg({1 \over x \bar x} -4\bigg)
\end{equation}
 has nonzero matrix elements  $\langle n_1 |V(x,\vec k_\perp) | n_2\rangle $ for all $n_1,n_2$ pairs. 
The perturbative part $V_{perp}$ for heavy quarks is the Coulomb term, with 
running coupling and other radiative corrections. Finally, the last term $ V_{spin} $
contains matrices in spin variables and in orbital momenta, which we will consider later.
 
\begin{figure}[t]
\begin{center}
\includegraphics[width=6cm]{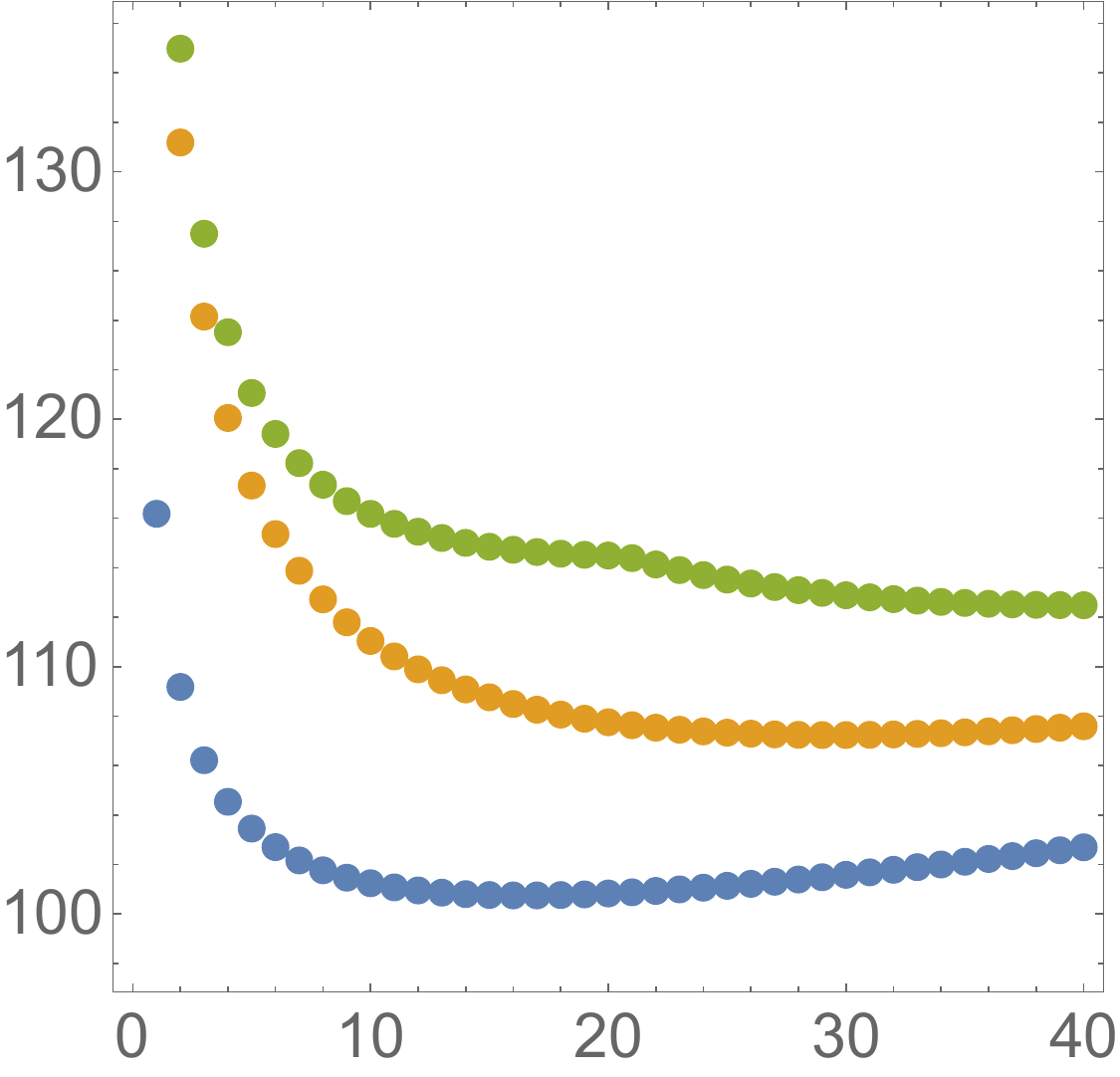}
\includegraphics[width=6cm]{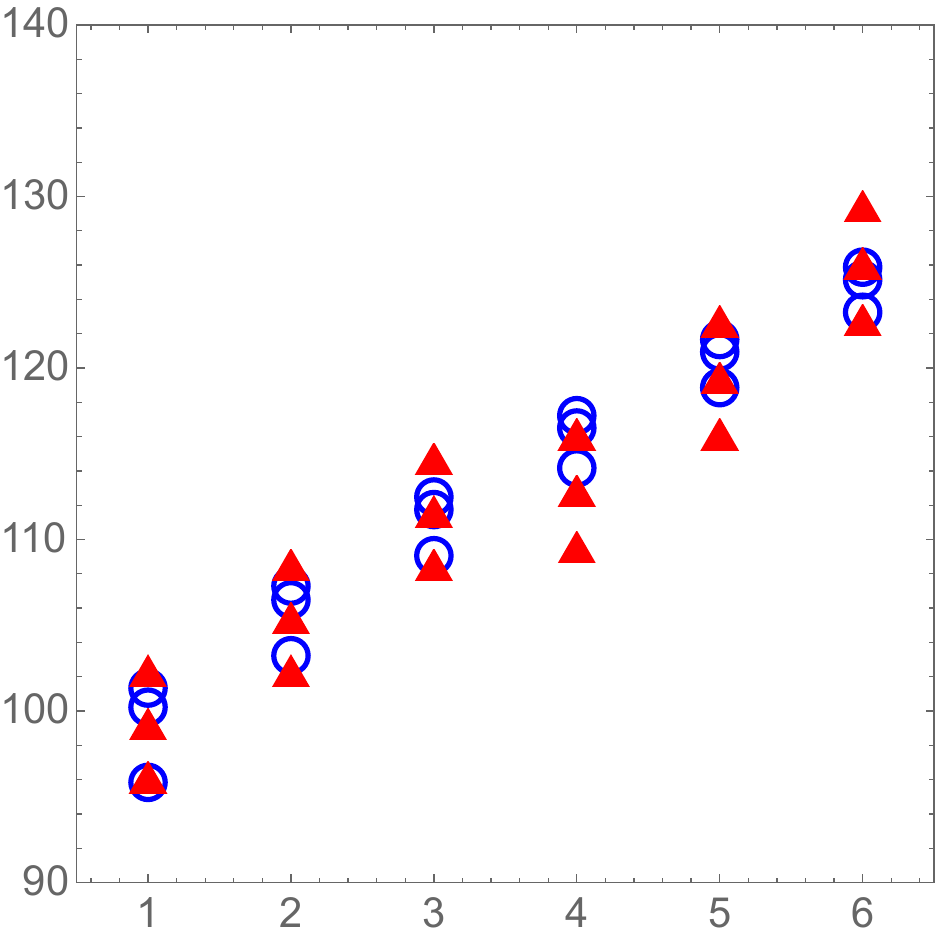}
\caption{Top: Squared masses $M^2_{n}$ for  $\bar b b$ mesons for $n=1,2,3$ versus the variational parameter $a$.
Bottom:  Squared masses  for $n=0..5$ (left to right) and orbital momentum 
$m=0,1,2$ (down to up), calculated from the light front Hamiltonian $H_{LF}$ (red triangles),
and shifted by a constant, $M^2_{n+1}-5\, GeV^2$. For comparison, the blue-circles show the squared masses $M^2_{n+1}$ calculated from Schroedinger equation in the CM frame,  with only  linear plus centrifugal potentials.}
\label{fig_M2_vs_a}
\end{center}
\end{figure}

We truncate the basis set to a 12$\times$ 12 matrix, and diagonalize $H_0+\tilde V$, to find its eigenvalues as a function of
the remaining parameter $a$. The results for the three lowest states 
$n=1,2,3$ are shown in Fig. \ref{fig_M2_vs_a} (top).  We see that while the minima in $a$
exist, they are not at the same value. Thus the dilemma: one can either select different
$a$ for different states, and then somehow re-orthogonalize them, or one can use
some ``optimal" value of $a$ common for all states, and then be sure that all
states are orthogonal.  Since the dependence on $a$ is rather flat, we opt  for
the second approach and use  $a=25$. 

The  calculated masses (shifted by a constant  ``mass renormalization", to make $n=0,m=0$ states the same) are shown in  Fig.\ref{fig_M2_vs_a}.
The bottom part shows  good agreement between the masses  obtained solving  the Shroedinger equation 
in the rest frame (blue circles),  and the masses following from the  light-front frame (red-triangles).
The slope is correct, and is determined by the same string tension $\sigma_T$. The splittings in orbital momentum are of the same scale, but not identical.
This is expected, as we compare the 2-dimensional  $m$-states on the light front,  with the 3-dimensional  $L$-states in the center of mass framew.

The  irregularity between the third and  fourth set of states, is due to our use of a 
 modest  basis set, with only three radial functions  (altogether 12 functions if one counts them with 4 longitudinal harmonics). 
 This can be eliminated using a larger set.

As a final note in this section, we recall that the chief goal of these calculations
is to generate the pertinent  LFWFs for all these states,  from the light front Hamiltonian  $H_{LF}$.  
The details about  the setting and some of these wavefunctions  can be found
in Appendices~\ref{2d_basis},\ref{bottom_LF},\ref{wf_m}.

\subsection{Matrix elements of  the Coulomb term and further operators on the LF}

We start with a calculation of the contribution of the Coulomb force, which demonstrate how to deal with  any function of transverse coordinate $F(r_\perp)$. 
Of course, the standard way is  to transform all functions into the coordinate representation.

Recall that our LFWFs  are defined using a transverse  oscillator in the {\em momentum representation}, 
and so one possible strategy is to trade
 $$\vec r_\perp \rightarrow i {\partial \over \partial \vec k_\perp} $$
as  in the confining potential. It can work for other polynomial 
functions $F(r_\perp)$ or their  Taylor expansions.
Unfortunately, for a transverse Coulomb potential 

\be 
V_{\perp}=-{C_C \over r_\perp} 
\ee
this  strategy  does not work. A straightforward solution is to Fourier transform
the LFWFs to coordinate representation. 
Note  that the LFWFs are of the form 
$$\psi(p_\perp,x)=\sum_n \phi_n(p_\perp) sin(\pi n x),$$
as a result, the integration over $x$ in the matrix element
$\langle \psi | F(r_\perp) | \psi \rangle $  removes terms with $n_1\neq n_2$, 
and reduces to

\be  
\langle \psi | F(r_\perp) | \psi \rangle=\sum_n \int d^2 r_\perp |\tilde \phi_n(r_\perp)|^2 F(r_\perp)
\ee 
where tilde stands for the Fourier transform. 

In particular, the lowest state in our basis has a simple Gaussian
form $\phi_1(p_\perp)\sim {\rm exp}(-A p_\perp^2)$, and its Fourier transform is 
also a Gaussian $ \tilde \phi_1(r_\perp) \sim  {\rm exp}(-{r_\perp^2 \over 4A})$. 
The average Coulomb contribution to the squared mass  $M^2_\Upsilon$ is then found to be

\bea
&&-4 M_b \langle \psi_1 | {C_C \over r_\perp}  | \psi_1 \rangle\nonumber\\
&& =-4 M_b C_C\sqrt{\pi \over 2A } \approx -15 \, GeV^2
\eea 
 which approximately agrees with $\Delta M^2_\Upsilon\approx -17 \, GeV^2$ obtained from 
 the Schroedinger equation in the CM frame (and shown in Fig.\ref{fig_6upsilons}). With growing Upsilon number their sizes grow, which
 reduces the Coulomb contribution.

In general, the operators to be averaged  (e.g. spin-dependent potentials) depend
on the invariant distance $\xi_x$, which includes the  longitudinal distance with the  derivative $id/dx$. In this case the matrix
element should be calculated using  the eigenvalue $l$ decomposition of 
this derivative operator. For example, by approximating the ground state  
 LFWF as
 
\be
\phi_0 (x, p_\perp)\approx \bigg(\frac {2\alpha}\pi\bigg)^{\frac 12}\,e^{-\alpha p_\perp^2/2}\sum_{odd\,l} \varphi_l \,{\rm sin}(l\pi x)
\ee
with a simple Fourier transform $p_\perp \rightarrow b_\perp$

\be
\label{FOURIER}
\tilde\phi_0(x, b_\perp)\approx \bigg(\frac 2{\pi\alpha}\bigg)^{\frac 12}\,e^{-b_\perp^2/2\alpha}\sum_{odd\, l} \varphi_l \,{\rm sin}(l\pi x)
\ee
one can use it to evaluate  matrix elements of a the  potential depending on this invariant $V(\xi)$
as follows

\begin{widetext}
\be
\label{FOURX}
\langle \phi_0|V(\xi_x)|\phi_o\rangle=
\int db_\perp\frac{e^{-b_\perp^2/\alpha}}{\pi\alpha}\sum_{odd_l}|\varphi_l|^2\,V\bigg(\bigg(\bigg(\frac{l\pi}{M\rho}\bigg)^2+\frac{b_\perp^2}{\rho^2}\bigg)^{\frac 12}\bigg)
\ee
\end{widetext}
The same procedure applies for the excited states. (\ref{FOURX}) shows that the natural transverse cutoff in $b_\perp\sim \pi/M\sim \pi/2m_Q$
for the heavy states.

\section{Spin and orbital momentum mixing of the LFWFs}~\label{sec_V}

In so far, we only discussed the LFWFs   diagonal in longitudinal
orbital momentum $L_z=m$. In general, this is not   a conserved quantity, but
for heavy quarkonia it is approximately conserved, as the spin and orbital momentum-dependent 
effects are suppressed by large quark masses. As we will proceed to light quark
states, this approach would become invalid, and  spin-spin and spin-orbit mixing 
is mandatory. 

\subsection{Parity on the light front}

There are some obvious differences between the description in the rest frame, and on the light front.
For instance, there are different symmetries:  3-dimensional  angular momenta $\vec S,\vec L,\vec J$ 
are reduced to their 2-dimensional transverse parts, 
with spin $\vec S$ and orbital momentum $\vec L$ projected onto 
longitudinal momentum $\vec P$. The projection of $J$ is denoted by  ``meson helicity" $\Lambda$. Obviously,
hadron states with different  $\Lambda$ values,
are treated  differently: say $\rho(\Lambda=0)$ and $\rho(\Lambda=\pm 1)$ have different wave functions 
(even more than one: see below). While masses, magnetic and quadrupole moments, etc should turned out to be the same, the 3-dimensional
rotation is some complicated transformation,  involving all components of the wave functions, and we will not attempt to explicitly use it. 

(This situation is of course not new. For example, different isospin components are also treated differently
in the rest frame. The $\bar d u$ charged states are described by a potential, while  the $\bar d d, \bar u u$ 
follow from annihilation. In the isospin symmetric limit, the  same mass for $\pi^+$ and $\pi^0$ needs to be explicitly demonstrated.)

Another important difference between  the rest frame and the light front frame notations, 
relates to the different definitions of parity. The usual P-parity 
is the sign change of $all$ 3 spatial coordinates, or mirror reflection. On the  light front,
 one would like to keep the main beam direction (of $\vec P$ ) intact,
so P is supplemented by  an additional rotation, by $\pi$ around some transverse axes: this operator is called $\hat Y$. The state's helicity $\Lambda$ changes sign, so its action for $\Lambda\neq 0$ is given by the so called Jacob-Wick
relation

\begin{equation} \hat Y | \vec P, \Lambda> =
	(-)^{S-\Lambda}\eta | \vec P, -\Lambda>
	\end{equation}
where $\eta$ is the intrinsic parity of the	state,
negative for quark-antiquark states, positive  for quark states, and negative for an antiquark plus gluon. 
Since parton's momenta are generically not in the direction of $\vec P$, $k_\perp \neq 0$, one should remember that only $one$ component of 
 $k_\perp$ changes sign under $\hat Y$. For $\Lambda=0$,  $\hat Y$ turns the state to itself, so for these states one can define $Y$-parity. 
 The changed definition of parity completely changes the parity mixing rules, and respectively the number of light front  wave functions, as described in detail in 
 \cite{Ji:2003yj}.

 Yet  the light front  wave functions in the helicity basis, have  different rules. 
 The classification is not done via the total $S_1,S_2,L$: only their $z$-components (meson directed)
 are used. They satisfy the obvious constraint
 $$ \Lambda=S_1^z +S_2^z +L^z $$
 In the following we will drop the $z$ upper-script.
 The $\Lambda=0$ states are eigenstates of $\hat Y$,
 minus for pions and plus for rho mesons:
 those have two wave functions
 (see section \ref{wf_m}) unlike $\Lambda=\pm 1$ states.  
 
 In~\cite{Shuryak:2021hng} we focused on the spin-dependent forces, with $\vec S_1\cdot \vec S_2,
 \vec S\cdot \vec L$ and tensor. Now we will have their analogues in beam projections,
 which we refer to by the same labels,  without vectors. Some of them are non-diagonal. For instance,  the tensor
force can mix $|S_1=S_2=\frac 12,L=0\rangle$ with  $|S_1=S_2=-\frac 12,L=2\rangle$.

\subsection{General form of LFWFs for  mesons with different $\Lambda$} 

To proceed with the spin effects, we need to get the full spin-orbit structure of the LFWFs. In order to
explain what we mean, consider a meson with total $helicity$ $\Lambda$ (with
longitudinal projection $J_z$). The total of two quark spins
$\vec S=\vec S_1+\vec S_2$  can be $S=0$ or $S=1$:
in the former case $\Lambda=L_z$, and in the latter there are three cases:
$L_z=\Lambda-1,L_z=\Lambda,L_z=\Lambda+1$. 
These 4 states (like e.g. $\chi_b,h_b$) are in general mixed by spin-dependent forces. Schematically
for the last three states,  the mixing  is captured by a 3x3 matrix  (the index of $\psi$ is the longitudinal
projection of orbital momentum, or $M_L$)  

\begin{widetext}
\bea \label{eqn_3by3}
H_{\Lambda}=
\begin{pmatrix} 
\psi_{\Lambda-1} , 
\psi_{\Lambda} ,
\psi_{\Lambda+1} 
\end{pmatrix}
\begin{pmatrix} 
	V_{diag} & V_{\pm 1} & V_{\pm 2} \\
	V_{\pm 1}  &  V_{diag} & V_{\pm 1}  \\
	V_{\pm 2} & V_{\pm 1}  & V_{diag}  
	\end{pmatrix}
\begin{pmatrix} 
\psi_{\Lambda-1} \\
\psi_{\Lambda} \\
\psi_{\Lambda+1} 
\end{pmatrix}
\eea
\end{widetext}
The spin-orbit $ V_{SL}$ interaction changes $L$ by $\pm 1$, and  the tensor interaction  $V_T$
changes $L$ by $\pm 2$ . So, in general, any meson has {\em three wave functions}, mixed by
spin- and $L_z=m$-flipping forces.

For  the important case of $\Lambda=0$  -- the pseudoscalar ($\eta_b...\pi$) and vector ($\Upsilon ...\rho$)  -- is now diagonal. Thus these
 mesons have different parity.  Furthermore, the  two additional components involved are 
$\Lambda=\pm 1$, which by symmetry are the same (up to different factors ${\rm exp}(\pm i\phi)$). So 
  (as derived in ~\cite{Ji:2003yj}), one needs only $two$  wave functions. 
  Yet for consistency,  for $\Lambda=0$ we still define
  their  wavefunctions with three components
 
 \begin{widetext}
\bea
 \label{WFPS}
	| P \rangle& =&\int d[1]d[2]{\delta_{ij} \over \sqrt{N_c}} \ \big[ 
	\psi_0^{P}(x,k_\perp) 
	\big( Q_{i\uparrow}^\dagger (1) \bar Q_{j\downarrow}^\dagger(2)- 
	Q_{i\downarrow}^\dagger(1) \bar Q_{j\uparrow}^\dagger (2)\big) \nonumber \\
&&\qquad\qquad\qquad +
	 \psi_{-1}^{P}(x,\vec k_\perp) 
	Q_{i\uparrow}^\dagger  (1) \bar Q_{j\uparrow}^\dagger(2)+ 
		\psi_{+1}^{P}(x,\vec k_\perp)  Q_{i\downarrow}^\dagger(1) \bar Q_{j\downarrow}^\dagger (2)
	\big] |0\rangle
	\eea
	\bea
 \label{WFVX}
	| V \rangle& =&\int d[1]d[2]{\delta_{ij} \over \sqrt{N_c}} \ \big[ 
	\psi_0^V(x,k_\perp) 
	\big( Q_{i\uparrow}^\dagger (1) \bar Q_{j\downarrow}^\dagger(2) +
	Q_{i\downarrow}^\dagger(1) \bar Q_{j\uparrow}^\dagger (2)\big) \nonumber \\
&&\qquad\qquad\qquad +
	\psi_{-1}^V(x,\vec k_\perp)Q_{i\uparrow}^\dagger  (1) \bar Q_{j\uparrow}^\dagger(2) -
		\psi_{+1}^V(x,\vec k_\perp) Q_{i\downarrow}^\dagger(1) \bar Q_{j\downarrow}^\dagger (2)
	\big] |0\rangle
\eea
\end{widetext}
with $N_c=3$. The subscripts $0$ and $\pm 1$ on the wave-functions, refer to  $L_z$, the z-projections of the orbital momentum. Note that compared to the notations in \cite{Ji:2003yj},
there are no explicit factors of $k_\perp^{\pm}=k_1\pm ik_2$ here because
they naturally belong to our wave functions,  consistently defined not only for $m=L_z=1$,  but for any $m$ value.

The  $\Lambda=0$ state of the vector mesons are called ``transversely polarized". 
The two other
polarizations, with  $\Lambda=\pm 1$ are ``longitudinally" polarized. They  are a bit more complicated, with
three components each with different wave-functions, corresponding to $L=2,1,0$. 

The invariant measure in (\ref{WFPS}-\ref{WFVX}) refers to the   on-shell covariant one,  with 
overall momentum conservation

\begin{eqnarray}
 d[1] d[2]= \frac{dx}{\sqrt{4x\bar x}}\frac{dp_\perp}{(2\pi)^3}
\end{eqnarray}
 Here $x, \bar x$ are the fraction of longitudinal momenta carried by particle-1 and the anti-particle-2, or
 $x=p^+_{1}/P^+$ and  $\bar x=p_2^+/P^+$ with $x+\bar x=1$.  The creation  and annihilation operators in (\ref{WFPS}-\ref{WFVX})
 obey  the  anti-commutation rules

\bea
 &&[Q_\alpha(k_1), Q_\beta^\dagger(k_2)]_+=\nonumber\\
&&\delta_{\alpha\beta}2k_1^+(2\pi)^3\delta(k_1^+-k_2^+)\delta(k_{1\perp}-k_{2\perp})\nonumber
\eea 
for equal light-front time, so that the $|P,V\rangle$ states are covariantly normalized on the light front, e.g.

\bea
\langle P|P^\prime\rangle =2P^+ (2\pi)^3\delta(P^+-P^{+\prime})\delta(P_\perp-P_\perp^\prime)\nonumber\\
\eea
It is readily checked that
the light front wave-functions in (\ref{WFPS}-\ref{WFVX}) are normalized by

\begin{equation}  
\label{NORMX}
\int {d^2 k_\perp dx \over (2\pi)^3} \big( |\psi_0|^2+   |\psi_1|^2 +  |\psi_{-1}|^2\big) =1
\end{equation}
Below we show that $\psi_0$ refers to the twist-2 and $\psi_{\pm 1}$ to the (tensor) twist-3 contribution
to the mesonic distribution amplitude.

\section{Quadrupole moment of vector mesons and $m\pm 2$ ``tensor" mixing }~\label{sec_VI}

To explain why the effects mixing spin and orbital momenta are important, let us take the 
classic example of  the {\em quadrupole moment}.
In the rest frame,  these  phenomena are well known in nuclear physics, for example the deuteron
$d=pn$ state has total $J=1$ and, in nonrelativistic notation,  it is  a mixture of  $L=0,J=1$
and $L=2,J=1$ states induced by the tensor force. 

In the light front formulation,  the rotational symmetry turns to  a {\em hidden symmetry},  
with apparent distinctions between longitudinal and transverse coordinates. Therefore, 
the LFWF mixing related to the quadrupole moment, takes two different forms:

\begin{enumerate}
\item for $\Lambda=1$ it is mixing of $\Psi_{0,0},\Psi_{0,2}$;
\item  for  $\Lambda=0$   it is mixing of $\Psi_{0,-1},\Psi_{0,1}$.
\end{enumerate}
Note that the indices here are the quantum numbers $n$ and $m$.

\subsection{S-D mixing  in the rest frame}

To assess the S-D mixing for Upsilon  in the center of mass frame, we need to first 
consider the splitting due to  the repulsive  centrifugal potential $(6/r^2)$  originating 
from the free Laplacian plus the Cornell potential, with the result
$$E_2-E_0=0.46669-(-0.47682)=0.943 \, GeV$$ 
(which should is still subject to corrections by spin-dependent forces). This value
is to be  compared to the empirical  mass difference
$$ M_{\Upsilon 2}-M_{\eta_b}=10.2325-9.3987=0.834 \, (GeV).$$

The S-D mixing requires two states with the same  $J=1$, which are constructed
in a standard way, via Clebsch-Gordon coefficients 

\bea
 \psi_0^{M_J=1} &=&\psi_0(r) Y^0_0 \chi^1_1  \nonumber \\
\psi_2^{M_J=1}& =&\psi_2(r)\bigg(\sqrt{3\over 5} Y^2_2 \chi^{-1}_1\nonumber\\
&&
-\sqrt{3\over 10} Y^1_2 \chi^0_1 + \sqrt{1 \over 10}Y^0_2 \chi^1_1\bigg)
\eea
Here  $Y^m_L(\theta,\phi)$ are spherical harmonics,  and $\chi^{M_s}_S$ are
states of total spin composed of $Q$ and $\bar Q$.  To proceed, we use  the standard notation for the tensor
force $V_T(r)S_{12} $,  and the non-diagonal matrix element with the angular integral

\bea
&& \int d\Omega  (Y^0_0 \chi^1_1) S_{12}  \nonumber\\
 &&\times \bigg(\sqrt{3\over 5} Y^2_2 \chi^{-1}_1-\sqrt{3\over 10} Y^1_2 \chi^0_1 + {1 \over \sqrt{10}}Y^0_2 \chi^1_1\bigg)=\sqrt{8}\nonumber\\
\eea
As a result, the  quadrupole moment is  given by an integral

\bea
 \langle Q \rangle_\Upsilon &=&{\epsilon_{02} \over 5\sqrt{2}} \int dr r^4 \Psi_{00}(r) \Psi_{02}(r)
 \nonumber\\ & \approx& \epsilon_{02} (0.14 \, GeV^{-2})
 \eea
where the admixing amplitude of the D state is $ \epsilon_{02}$ . If we assume it to be small, it is then given by the 
perturbative  matrix element of the tensor mixing operator,  sandwiched between  states calculated using the Cornell potential

\bea
  \epsilon_{02}={ \sqrt{8} \int dr r^2  \Psi_{00}(r) V_{\pm 2}(r)  \Psi_{02}(r) \over E_2-E_0}
\eea
For an estimate, we may use the perturbative  contribution  with

\bea
V_T(r)=\frac 43 \frac{3\alpha_s}{r^3}
\eea


\subsection{Quadrupole moment of Upsilon meson from the light front Hamiltonian }

In this subsection we still 
consider  the case of $\Upsilon$, a vector meson made of $
\bar b b$ quark pair. In the rest frame  frame we just discussed, the cases of transversely polarized 
$\Lambda=0$ and longitudinally polarized $\Lambda=\pm 1$ are related by the $O(3)$
rotational symmetry. The matrix elements of the various tensor operators over the corresponding states,
are tied by the  Wigner-Eckart theorem, and given by Clebsch-Gordon coefficients times a
``reduced" (rotationally invariant) matrix element independent of the meson orientation.

\begin{figure}[t]
\begin{center}
\includegraphics[width=6cm]{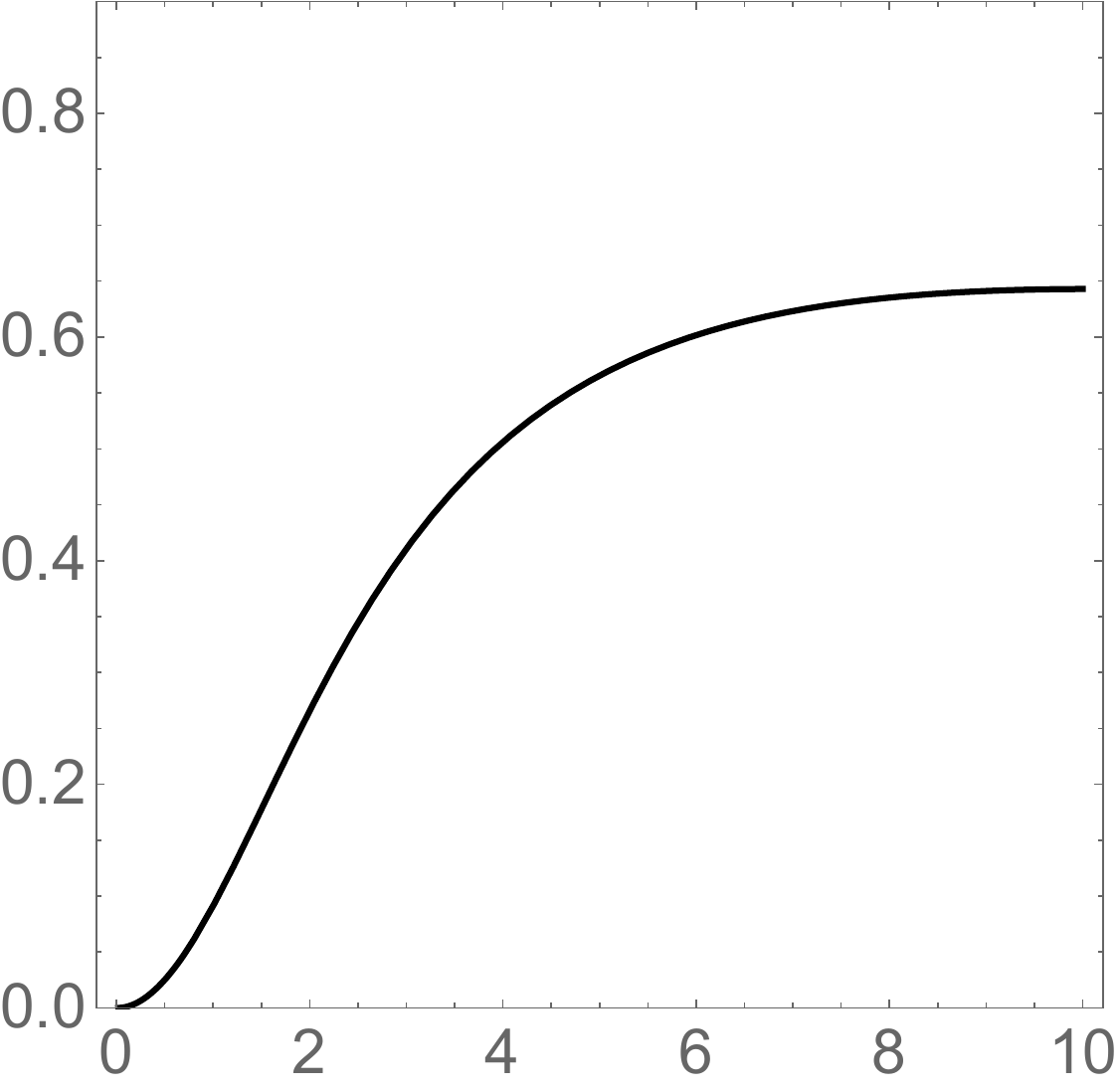}
\includegraphics[width=6cm]{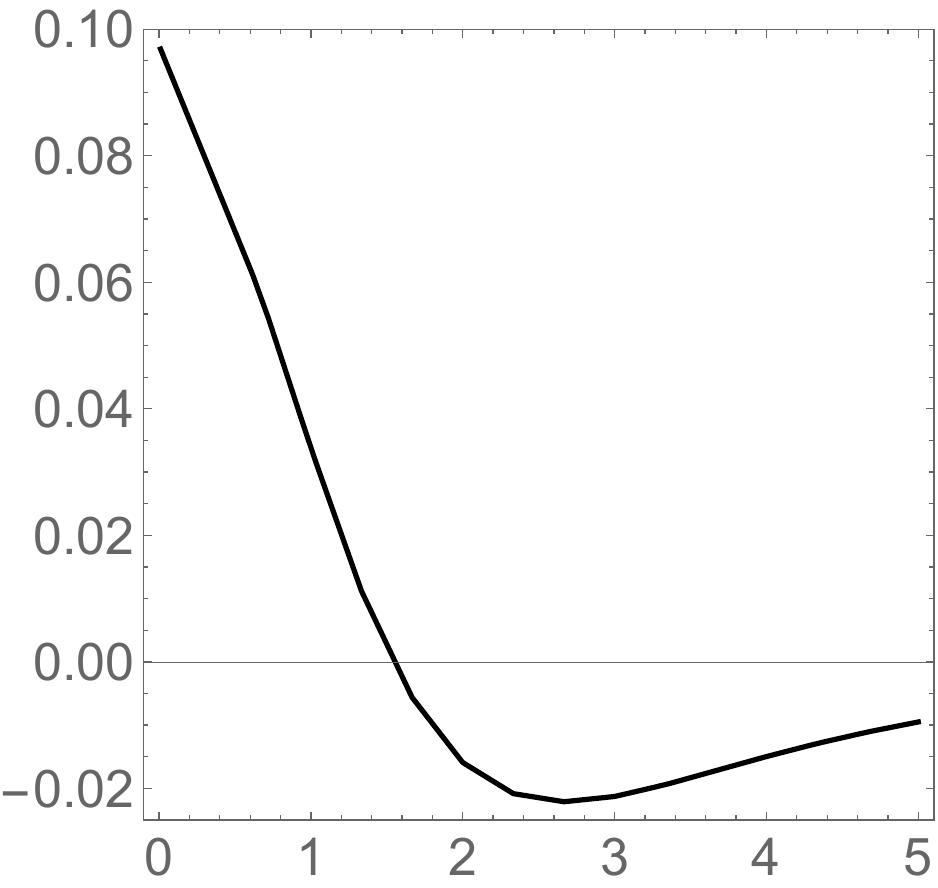}
\caption{The central part of the instanton induced potential $\mathbb V_C(\xi_x)$ versus the distance $\xi_x=r$ (top),
and its second derivative $\mathbb V_C^{\prime\prime}(\xi_x)$ (bottom). See text.}
\label{fig_Vc_VT}
\end{center}
\end{figure}

In the light front Hamiltonian $H_{LF}$ in (\ref{HLFX}), the  spin-tensor potential  for heavy quarkonia
comes from the last term with the  instanton-induced effects (\ref{SPINFULL}). 
The spin operator can be re-written in a more transparent way as 

\bea
&& \bigg(\sigma_{1\perp}\cdot\hat b_{21}\sigma_{2\perp}\cdot\hat b_{21}-\frac 12\sigma_{1\perp}\cdot\sigma_{2\perp}\bigg)\nonumber\\
&&={1\over 4} \big(\sigma_{1-}\sigma_{2-}e^{2i\phi}+
\sigma_{1+}\sigma_{2+}e^{-2i\phi}\big)
\eea
The dependence on the azimuthal angle $\phi$ reflects on the  mixing
 between the $m$ and $m\pm 2$ states. We recall that the  instanton contribution
to the central potential was discussed in detail in \cite{Shuryak:2021hng}, for the  ``dense instanton ensemble" with
diluteness parameter $\kappa$ set to one. 
The plot of the central potential $\mathbb V_C(\xi_x)$ and its second derivative $\mathbb V_C^{\prime\prime}(\xi_x)$ are given in 
Fig.\ref{fig_Vc_VT}. Note the change in  sign at $b_\perp\sim 1.5 \, GeV^{-1}$,
a  distance comparable to the size of Upsilon.

With our usual
approximation for heavy quarkonia, $M\approx 2 m_Q$ and $\xi_x\approx b_\perp$,
its contribution to the mixing part of $H_{LF}$ takes the form

\bea
&&  \langle n_1 m | V_{\pm 2} | n_2,m\pm 2 \rangle \nonumber \\
&&=  \int d^2 b_\perp dx \Psi_{n_1m}^*
V_T(b_\perp) e^{-2i\phi}  \Psi_{n_2m+2} 
 \big] \nonumber\\
\eea
We show the factors depending on $\phi$ explicitly,  but omit the spin operators.

To simplify the wave functions,
let us for now ignore $\tilde V$ in the Hamiltonian, which means using
instead of $\Psi_{0,0},\Psi_{0,2}$ LFWFs,  the functions $\psi_{0,0},\psi_{0,2}$
of the oscillator basis, (\ref{eqn_psi_n0}) and (\ref{eqn_psi_n2}). Recall  that
in the corrdinate representation for the $\bar b b$ mesons,  the size parameter $\beta\approx 0.62 \, GeV$. The result is 

\be  
\label{V12X}
\langle 0 0 | V_{\pm 2} | 0, 2 \rangle\approx -0.011 \, GeV^2
\ee
Note that if the size integral is split into the contributions stemming from
the small plus large radial intervals, i.e. $[0,1.5 \, GeV^{-1}]$ plus
$[1.5 \, GeV^{-1} ,\infty]$, we find 0.007 and -0.018, respectively.

When (\ref{V12X}) is divided by the difference of the mass squared for the 
two mixing states

\bea
\Delta M^2&=&M_{\Upsilon 2}^2-M_{\eta_b}^2\nonumber\\
&=&10.2325^2-9.3987^2\approx 16.4 \, GeV^2\nonumber
\eea
we get our  estimates of the mixing parameter
$ \epsilon_{02}\approx 0.00064$. As a result, 
the estimate for the Upsilon quadrupole moment is then

\be 
Q_\Upsilon\sim  2\epsilon_{02} \int d^2b_\perp dx \Psi_{0,0} \Psi_{0,2} b^2_\perp  \approx -0.0095\, \, GeV^2 
\ee
which means that the usually quoted combination is

\be   
Q_\Upsilon M_\Upsilon^2 \approx - 0.87
 \ee

As we will see in the next subsection, it is right in the ballpark of
other determination.
Unfortunately, this result is relatively uncertain since it comes from
significant cancellations of small and large ranges in the  $b_\perp$-mixing
integral. Deformations of the instantons -- e.g. in instanton-antiinstanton ``molecule"
configuration described by streamline or thimble configuration -- would change
this number. Putting this observation into a positive direction, we may 
conclude that the quadrupole moments
of mesons,  are sensitive to the exact nature of the nonperturbative vacuum fluctuations.

We recall that in~\cite{Shuryak:2021fsu},  we  extracted  the matrix element
of the tensor force from the masses of the  $P$-wave states of mesons with different quark species, ranging from the heavy $\chi_b$
to the light $K, \pi$. We noted  that this matrix element  changes sign, in going from heavy to  light mesons.   This  observation is consistent with the calculation of this section. This  issue clearly deserves further studies.

\subsection{Quadrupole moments of vector mesons from lattice and other approaches}

There have been several lattice measurements of the quadrupole moments 
of vector mesons, and in all fairness we will not be able to cover them in this 
comparative study. In a recent lattice study of vector mesons composed of light, 
strange and charmed quarks with $V=\rho,K^*,\rho_s,\rho_c$ 
(the latters carry artificial charge assignments),  it was numerically found  
that $Q_V M_V^2\approx -0.3$~\cite{Xu:2019ilh}, which is
comparable to an earlier study with $Q_V M_V^2\approx-0.23(2)$~\cite{Dudek:2006ej}. 
When extrapolated to bottomium, the recent lattice result gives
$Q_{\Upsilon}\approx -0.003\, GeV^2$,  with  a mixing
parameter $\epsilon_{02}\approx 0.02$ from our analysis.

Adhikari et al \cite{Adhikari:2018umb} have used their version of light-front Hamiltonian and wave functions,
 and calculated form-factors for the lowest states of charmonia and bottomonia. From their
Table V, we  see that their value for Upsilon is  larger $Q_\Upsilon M_\Upsilon^2=-0.731(9)$. 
In sum, the spread of these numbers is about a factor of 3, so the
magnitude of the quadrupole moment of Upsilon remains   relatively uncertain.

\section{General spin and orbit mixing for   light quarks}~\label{sec_VII} 

In so far, we have considered mixing between the $m=0$ and $m=2$ components
of the quarkonia wave function by the tensor force $V_{\pm 2}$.  Simarly, we 
can   include spin-orbit force $V_{\pm 1}$ and spin-spin forces,
generating the whole $3\times 3$ mixing matrix. However, since we know that
all mixing is suppressed by powers of the heavy quark mass, we  can treat
these mixings perturbatively and additively, as we did above.

Instead, we now switch to the more involved case of light quarks,
where the mixing is not expected to be suppressed. Of course, this is well known from the spectroscopy
in the rest frame: heavy quarkonia are nonrelativistic, while
light quark systems are not. In the rest frame, it is difficult to compare these two limiting cases
of the meson spectroscopy. Fortunately, in the light front the comparison is possible, as light and  heavy
quarks are treated democratically. 
The light front  Hamiltonian  has the same form for both cases, with only few
parameters  due for change. The only special case
is the pion as a Goldstone mode,  that we will address in the next section
(see also our qualitative analysis in~\cite{Shuryak:2021hng}).

 The  ``basic problem" of two constituent quarks connected by a confining string
 was already considered in~\cite{Shuryak:2021hng}. 
 There we did not yet have mixing of states with different (transverse)  orbital momentum $m$, and considered 
only the set of functions with  $m=0$. The basis functions with $m\neq 0$ are
discussed in Appendix~\ref{wf_m}, including the transition from the momentum to the coordinate representation.

The general form of the mixing matrix $H_\Lambda$ for a meson with helicity $\Lambda=J_z$,
 was already given in~(\ref{eqn_3by3}). The derivation of the perturbative and instanton-induced 
spin- and orbital-$m$-changing effects were given in our recent analysis
in~\cite{Shuryak:2021hng} and above. Our current task is to evaluate the corresponding matrix elements. 

The diagonal in the $m$-part or $V_{diag}$, consists of two parts, the one coming from
the spinless $H_0+\tilde V$ Hamiltonian, and the one from the spin part. There is no need to
describe in details the  former. In brief, the values of the squared masses  for the $m=0,1,2$
states were obtained after its diagonalization. Using the states detailed  in the Appendix, we get
for the longitudinally polarized vector $\Lambda=1$ case the following diagonal elements
\be 
H_1^{00}=2.229, \,  H_1^{11}=2.833,\,  H_1^{22}= 3.434\, (GeV^2)  
\ee
Three  spin states  $S=1,S_z=1,0,-1$  are
 $$| \uparrow \uparrow\rangle,  (| \uparrow \downarrow\rangle +| \downarrow \uparrow\rangle )/\sqrt{2}   ,| \downarrow \downarrow\rangle $$
respectively. The
spin-spin forces are proportional to the same value
$\langle \vec S_1 \cdot\vec S_2 \rangle=1/4$, and the corresponding matrix elements of the perturbative (\ref{PERP1}), and
instanton-induced  (\ref{eqn_inst_spin_orb})  potentials with $|\Psi_{0m}|^2$.

\begin{figure}[h!]
\begin{center}
\includegraphics[width=6cm]{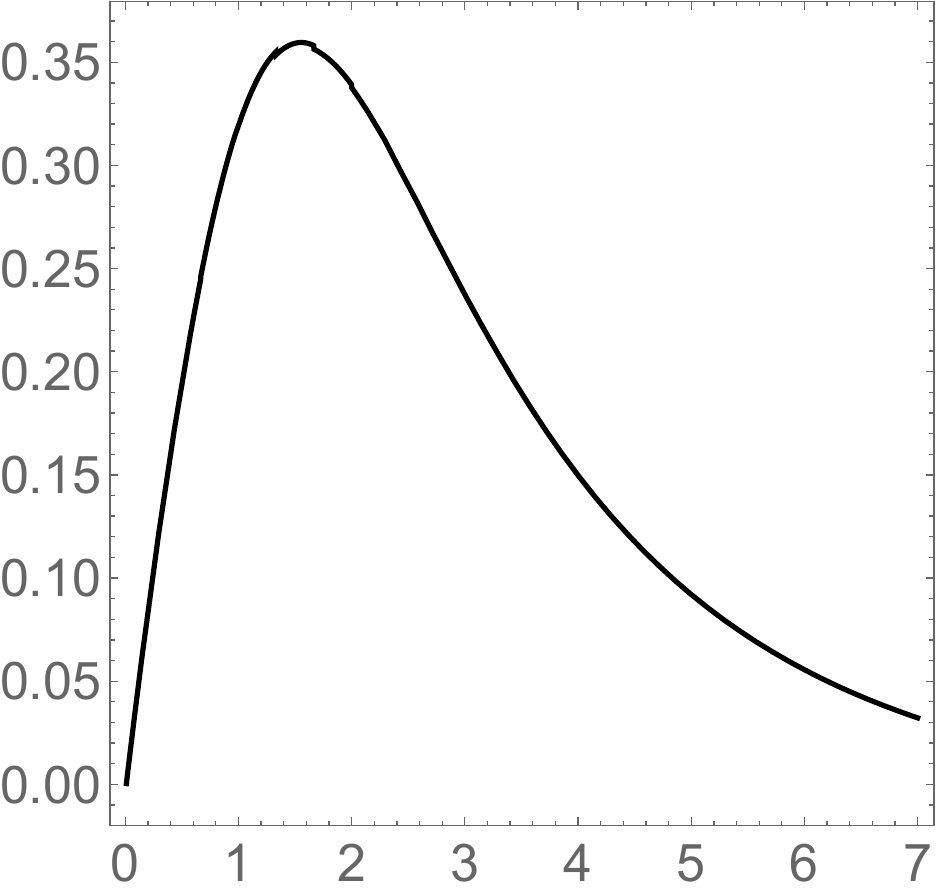}
\caption{Instanton induced spin-orbit potential on the light front $\mathbb V_{C}^\prime(\xi_x)$. See text. }
\label{fig_spin_orbit_inst}
\end{center}
\end{figure}

The near-diagonal $V_{\pm 1} $ part of the mixing matrix is due to the spin-orbit forces, 
from the perturbative  (\ref{PERP1}) and instanton-induced effects (\ref{eqn_inst_spin_orb}). The   former is  proportional to $-\hat S_1 \cdot \hat L_2+\hat S_2 \cdot \hat L_1$, in which $\hat L_2=-\hat L_1$ so the two terms are added into the total spin $\hat S$. We only consider the  non-diagonal operators $S^+ L^-+ S^- L^+$  that flip the spin and $m$ by $\pm 1$.
The perturbative potential is   $\sim 1/\xi_x^3$. If  we ignore  the longitudinal distance and
use $\xi_x^2\approx \vec b_\perp^2$,  we  may worry of  the convergence of the integrand at the origin.
In the transition between $m=0$ and $m=1$, one indeed finds a logarithmic divergence
\bea
&&\int d^2b_\perp  \Psi_{00} {1 \over b^3}\Psi_{01 }\nonumber\\
&& \sim \int db_\perp  b_\perp *{1 \over b_\perp^3}*b_\perp  \sim log(b_{min})
\eea
since at small $b_\perp$, $\Psi_{00}\sim b_\perp^0,\Psi_{01 }\sim b_\perp$. This logarithmic divergence is cut-off by the small 
longitudinal distance of about $\pi/M\approx \pi/2m_Q$. In contrast, 
in the transition between $m=1$ and $m=2$ the integral has  $\Psi_{02}\sim b_\perp^2$ instead of $\Psi_{00}$, so it is convergent.

The instanton-induced spin-orbit contributes to the 
 light front Hamiltonian $H_{LF}$, with the corresponding potential $V_{C}^\prime(\xi_x)$ shown in~Fig.\ref{fig_spin_orbit_inst}. 
It is regular at the origin, but with a  relatively small range of about an  instanton size $\sim 1.5\, GeV^{-1}\sim 0.3 \, fm$.  

The $V_{\pm 2} $ or tensor forces were  discussed in details above, for the case of the transition between
the $m=0$ and $m=2$ states. The corresponding instanton-induced potential  is shown  in the lower Fig.\ref{fig_Vc_VT}.

After the evaluation of  all matrix elements, we obtain the following mixing matrix

\begin{widetext}
\bea \label{eqn_mixing}
H_{\Lambda=1} &=&
\begin{pmatrix} 
M_0^2 + C^{00} + SS^{00}_C + SS_{inst}^{00} , &
  SL_C^{01}+ SL_{inst}^{01} , &
  T_{inst}^{02} \\
   SL_C^{01}+ SL_{inst}^{01} , &  M_1^2 + C^{11} + SS_C^{11} + SS_{inst}^{11} , &
  SL_{pert}^{12} + SL_{inst}^{12}\\
   T_{inst}^{02}, &
  SL_{pert}^{12}+ SL_{inst}^{12} , & 
  M_2^2 + C^{22} + SS_c^{22} + SS_{inst}^{22} 
	\end{pmatrix}  \nonumber 
	\eea
	\bea
=\begin{pmatrix} 
2.06789 & 0.289719 & -0.100297 \\
0.289719 & 2.66282 & 
  0.0617886 \\
  -0.100297 & 0.0617886 & 3.2959
	\end{pmatrix}
\eea
\end{widetext}
All entries in  (\ref{eqn_mixing}) are explained in~Appendix~\ref{mixing_me}.
The mixing changes the squared masses of the three $\Lambda=1$ states as follows

\ba   
\{M_0^2, M_1^2,M_2^2\} &=&\{2.229,  2.833, 3.434  \} \rightarrow  \nonumber \\ \{M_a^2, M_b^2,M_c^2\} &=& \{1.940, 2.779, 3.307 \} \nonumber \ea
The largest change is, as expected, a downward shift of the ground state. 

One may think that the main shifts are due to Coulomb and spin-spin effects,
and that the mixing is  small. This is not the case, as one can see from
the mixing coefficients  associated to the wavefunctions for these  three states

\ba \Psi_a&=&0.922252 \psi_0 -0.377037\psi_1+ 0.0854116\psi_2 \nonumber \\
\Psi_b&=&0.38106 \psi_0 +0.923825\psi_1-0.0365531 \psi_2 \nonumber \\
\Psi_c&=&-0.0651235 \psi_0+0.0662586 \psi_1+0.995675\psi_2 \nonumber
\ea
There is significant 0-1 mixing due to spin-orbit interactions.

\section{Including t' Hooft effective Lagrangian}~\label{sec_THH}

The zero-mode contributions due to tunneling are captured by the $^\prime$t Hooft determinantal interaction, in the rest frame. The 3-flavor 
determinantal interaction reduced to 2-flavor reads~\cite{Shuryak:2021fsu,Shuryak:2021hng}

\bea
\label{THOOFT}
V_{TH}(1,2)=&&
-\frac 14 |\tilde\kappa_2| A_{2N}\bigg(1-\tau_1\cdot \tau_{2}\bigg)\nonumber\\
&&\times \bigg(1-16B_{2N} S_1\cdot S_{2}\bigg)\,\delta(\vec{x}_{12})\nonumber\\
\eea
with the pair spatial distance $\vec{x}_{12}=\vec{x}_1-\vec{x}_{2}$, in the ultra-local approximation for the instanton.
However, the fermionic zero modes ride a non-local instanton tunneling process, and its interpretation in a boosted frame,
requires analytical continuation.

To derive the analogue of (\ref{THOOFT}) on the light front, 
we need to show how to analytically continue the fermionic tunneling process to the light front, where
the in-out quarks are nearly on mass-shell.
We need an LSZ reduction scheme in {\it Euclidean signature}, that extends to the
light front in Minkowski signature. For that, we follow the proposal suggested 
in~\cite{Liu:2021evw}.

\subsection{LSZ reduction in the rest frame}

For two flavors, the LSZ reduced $^\prime$t Hooft vertex between on-shell light quarks in the rest
frame is

\bea
\label{RF0}
\bigg<\bigg[\chi_R^\dagger (k_2)i k_2 \Phi_0(-k_2)\frac 1{im_q}\Phi_0^\dagger (k_1) i\bar{k}_1\chi_L(k_1)\bigg]\times  \nonumber\\
\bigg[\chi_R^\dagger (\underline{k}_2)i{\underline{k}}_2 \Phi_0(-\underline{k}_2)\frac 1{im_q}\Phi_0^\dagger (\underline{k}_1) i\bar{\underline{k}}_1\chi_L(\underline{k}_1)\bigg]\bigg>_U
  \nonumber \eea
The averaging in (\ref{RF0}) is over the $SU(N_c)$ color moduli $U$.
Here  each factor is on mass shell using the long time limit in  Euclidean signature.
More specifically, for the in-coming left-handed and on-shell $\chi_L(k_1)$ going through
an instanton, we define

\bea
\label{RF1}
&&\Phi_0^\dagger (k_1) ik_1\chi_L(k_1)\,e^{-|\vec k_1||T|}\nonumber\\
&&=\lim_{T\to\infty}\int d^3y \bigg(\frac{\rho^{\frac 32}}{\pi}\frac {U^\dagger\epsilon y\chi_L(k_1)}{y^4\Pi_y^{\frac 32}}\bigg)e^{-i\vec k_1\cdot \vec y}
\eea
with $\Pi_y=1+\rho^2/y^2$,  and  $y=(-T, \vec y)$.  In the large {\it Euclidean} time limit $\Pi_y\rightarrow 1$, and the y-integration reduces to

\bea
\label{RF2}
&&\lim_{T\to\infty}\int d^3y \,\frac{-T{\bf 1}-i\vec\sigma \cdot\vec y}{(T^2+\vec{y}^2)^2\Pi_y^{\frac 32}}\,e^{-i\vec k_1\cdot \vec y}\nonumber\\
&&=\pi^2({\bf 1}-\vec\sigma\cdot \hat{k}_1)\,e^{-|\vec k_1||T|}
\eea
Note the appearance of the mass-shell condition $E_1=|\vec k_1|$ in the exponent, supporting the LSZ reduction in Euclidean signature. As a result, (\ref{RF1})  simplifies

\begin{widetext}
\bea
\label{RF3}
\Phi_0^\dagger (k_1) ik_1\chi_L(k_1)=(\pi\rho^{\frac 32})\big[U^+\epsilon \,({\bf 1}-\vec\sigma\cdot \hat{k}_1)\,\chi_L(k_1)\big]=
(2\pi\rho^{\frac 32})\big[U^+\epsilon \,\chi_L(k_1)\big]
\eea
Since $\chi_L(k_1)$ is left-handed in the small current mass $m_q$ limit, the right-most result follows.
A repeat of this analysis, yields (\ref{RF0}) in the form

\bea
\label{RF4}
\bigg(\frac{4\pi^2\rho^3}{im_q}\bigg)^2\bigg<\big[\chi_R^\dagger (k_2)\epsilon U\big]\big[U^\dagger\epsilon\chi_L(k_1)\big]\times
\big[\chi_R^\dagger (\underline{k}_2)\epsilon U\big]\big[U^\dagger\epsilon\chi_L(\underline{k}_1)\big]\bigg>_U
\eea
\end{widetext}

\subsection{LSZ reduction on the light front}

To carry the preceding analysis to the light front, we first carry the analogue of the LSZ reduction along $y_+={\rm cos}\theta y_4+{\rm sin}\theta y_3$ for large $y_+$ in Euclidean signature,
integrate over the remaining orthogonal directions $y_-={\rm sin}\theta y_4-{\rm cos}\theta y_3$ and $y_\perp$,
and then analytically continue $\theta\rightarrow -i\chi$. More specifically,  the analogue of (\ref{RF2}) is now 

\begin{widetext}
\bea
\label{RF5}
&&\lim_{y_+\to\infty}\int dy_-dy_\perp  \,e^{-ik_{1-}y_--ik_{1\perp}\cdot y_\perp}\nonumber\\
&&\times\bigg(
\frac {[y_+({\rm cos}\theta-i{\rm sin}\theta \sigma_3)+y_-({\rm sin}\theta+i{\rm cos}\theta \sigma_3)-i\sigma_\perp\cdot y_\perp]}{(y_+^2+y_-^2+y_\perp^2)^2\Pi_y^{\frac 32}}\bigg)\nonumber\\
&&=\pi^2\bigg(({\rm cos}\theta{\bf 1}-i{\rm sin}\theta \sigma_3)-\frac{ik_{1-}}{k}({\rm sin}\theta{\bf 1}+i{\rm cos}\theta \sigma_3)-i\sigma_\perp\cdot y_\perp\bigg)\,e^{-k|y_+|}
\eea
\end{widetext}
in the large $y_+$ limit, with

\bea 
\label{RF6}
k_{1+}&=&{\rm cos}\theta k_4+{\rm  sin\theta}k_3\nonumber\\
k_{1-}&=&{\rm sin}\theta k_4-{\rm cos\theta}k_3\nonumber\\
k&=&(k_{1-}^2+k_\perp^2)^{\frac 12}
\eea
The analytical continuation $\theta\rightarrow -i\chi$, $y_4\rightarrow iy_0$ and $k_4\rightarrow -ik_0$ yield (\ref{RF5}) in the form

\bea
\label{REF6X}
2\gamma\pi^2({\bf 1}-\sigma_3)\,e^{- k|y_+|}\rightarrow \pi^2 ({\bf 1}-\sigma_3)\,e^{- k|y_+|}\nonumber\\
\eea
Modulo the overall factor $2\gamma=2P^+/M$ (to be absorbed in the 
new normalization of the states on the light front),
 (\ref{REF6X}) is compatible with (\ref{RF3}) in the large momentum limit.

In retrospect, (\ref{REF6X}) shows that (\ref{RF4}) extends to the light front  in the form

\begin{widetext}
\bea
\label{RF7}
\bigg(\frac{4\pi^2\rho^3}{im_q}\bigg)^2\bigg<\big[\chi_{3R}^\dagger (k_2)\epsilon U\big]\big[U^\dagger\epsilon\chi_{3L}(k_1)\big]\times
\big[\chi_{3R}^\dagger (\underline{k}_2)\epsilon U\big]\big[U^\dagger\epsilon\chi_{3L}(\underline{k}_1)\big]\bigg>_U
\eea
\end{widetext}
with  the $L,R$ polarizations solely along the 3-direction.  In other words, on the light front
 {\it helicity and chirality} are identified: a left-handed quark  with spin down, flips to a
right-handed quark with spin up as it tunnels through an instanton on the light front.
The opposite flip takes place through an anti-instanton.

\begin{widetext}

\subsection{ Zero mode induced ($^\prime$t Hooft Lagrangian)  interaction on the light front}

(\ref{RF7}) does not carry any  form factor in leading order, by the LSZ reduction. Its contribution to the invariant
meson  squared mass is

\bea
\label{RF8}
&&-\int\prod^2_{i=1}\frac{dk^+_i}{4\pi k_i^+}\frac{d\underline{k}^+_i}{4\pi \underline{k}_i^+}\frac{dk_{i\perp}}{(2\pi)^2}\frac{d\underline{k}_{i\perp}}{(2\pi)^2}\nonumber\\
&&\times
(2\pi)^3\,2P^+\delta\bigg(k_1^++\underline{k}_1^+-k^+_2-\underline{k}_2^+\bigg)\delta\bigg(k_{1\perp}+\underline{k}_{1\perp}-k_{2\perp}-\underline{k}_{2\perp})\bigg)\nonumber\\
&&\times\frac {n_{I+\bar I}}2 \bigg(\frac{4\pi^2\rho^3}{im_q}\bigg)^2\bigg<\big[\chi_{3R}^\dagger (k_2)\epsilon U\big]\big[U^\dagger\epsilon\chi_{3L}(k_1)\big]\times
\big[\chi_{3R}^\dagger (\underline{k}_2)\epsilon U\big]\big[U^\dagger\epsilon\chi_{3L}(\underline{k}_1)\big]+L\leftrightarrow R\bigg>_U\nonumber\\
\eea
with the anti-instanton contribution added. Recall that in the rest frame,  an effective form factor of the form (Euclidean signature)

\bea
\label{RF9}
\bigg(M(k_1)M(k_2)M(\underline{k}_1)M(\underline{k}_2)\bigg)^{\frac 12}\rightarrow 
\bigg(M(k_{1\perp})M(k_{2\perp})M(\underline{k}_{1\perp})M(\underline{k}_{2\perp})\bigg)^{\frac 12}\nonumber\\
\eea
is  induced, with $M(k)$ the running constituent mass. 
(\ref{RF9})  regulates the momentum transfers, since the in-out quark pair is not on mass-shell. 
We expect the rightmost form factor in  (\ref{RF9}) to carry to the light front
when the strict mass-shell limit is lifted as noted in~\cite{Kock:2020frx,Kock:2021spt}.

The color averaging in (\ref{RF8}) is similar to the color averaging carried in the rest frame (see Eq. 75 in~\cite{Shuryak:2021fsu}).
The result for the bracket in the mesonic channel  is

\bea
\label{TNL2}
&&\bigg[\bigg(\overline U(k_2, s_2) U(k_1, s_1)\overline V(k_4, s_4) V(k_3, s_3)+\overline U(k_2, s_2)\gamma^5U(k_1, s_1)\overline V(k_4, s_4) \gamma^5V(k_3, s_3)\nonumber\\
&&-\overline U(k_2, s_2)\tau^a U(k_1, s_1)\overline V(k_4, s_4) \tau^aV(k_3, s_3)-\overline U(k_2, s_2)\gamma^5\tau^aU(k_1, s_1)\overline V(k_4, s_4) \gamma^5\tau^aV(k_3, s_3)\nonumber\\
&&-4B_{2N}
\bigg(\overline U(k_2, s_2)\sigma^a U(k_1, s_1)\overline V(k_4, s_4) \sigma^aV(k_3, s_3)+\overline U(k_2, s_2)\gamma^5\sigma^aU(k_1, s_1)\overline V(k_4, s_4) \gamma^5\sigma^aV(k_3, s_3)
\nonumber\\
&&-\overline U(k_2, s_2)\sigma^a\tau^bU(k_1, s_1)\overline V(k_4, s_4) \sigma^a\tau^bV(k_3, s_3)-\overline U(k_2, s_2)\gamma^5\sigma^a\tau^bU(k_1, s_1)\overline V(k_4, s_4) \gamma^5\sigma^a\tau^bV(k_3, s_3\bigg)
\bigg]\nonumber\\
\eea
\end{widetext}
with $U(k,s),V(k,s)$ the quark and the antiquark LF spinors  respectively (see  Appendix~\ref{APP-LFX}).
Note that the same interaction holds for
$$(\overline UU)(\overline VV)\rightarrow (\overline UV)(\overline VU)\,.$$
through Fierz re-arrangements.

Besides the standard flavor-determinantal character of the squared mass operator (\ref{RF8}) after color averaging, 
its chief effect  is to flip the helicity/chirality/spin  of the in-out nearly on-shell light-front quark pair,  in the chiral limit.
For three light flavors $u,d,s$ in QCD, the arguments are similar but the strange quark loops in the sea as $\langle \bar s s\rangle<0$. The reduction to two flavors $u,d$
is structurally  identical to (\ref{RF8}), with only the overall coupling modified and {\it sign switched}. 

\begin{widetext}
On the LF, each of the contributions in  (\ref{TNL2}) is spin-valued. If we denote by $[s_2s_1]$ the entries with $s_1$
for the initial spin and $s_2$ for the final spin, then the matrix valued forms are
\bea
\label{UVX1}
\overline U_L(k_2, s_2) U_R(k_1, s_1) &=&+ \sqrt{k_1^+k_2^+}
\begin{pmatrix}
\frac{m_{Q_1}}{k_2^+}&0 \\
\frac{k_{1R}}{k_1^+}-\frac{k_{2R}}{k_2^+}& \frac{m_{Q_1}}{k_1^+}
\end{pmatrix}=+\sqrt{x_1x_2}
\begin{pmatrix}
\frac{m_{Q_1}}{x_2}&0 \\
\frac{k_{1R}}{x_1}-\frac{k_{2R}}{x_2}& \frac{m_{Q_1}}{x_1}
\end{pmatrix}
\nonumber\\
\overline U_R(k_2, s_2) U_L(k_1, s_1) &=& +\sqrt{k_1^+k_2^+}
\begin{pmatrix}
\frac{m_{Q_1}}{k_1^+}&\frac{k_{2L}}{k_2^+}-\frac{k_{1L}}{k_1^+} \\
0&\frac{m_{Q_1}}{k_2^+}
\end{pmatrix}=+\sqrt{x_1x_2}
\begin{pmatrix}
\frac{m_{Q_1}}{x_1}&\frac{k_{2L}}{x_2}-\frac{k_{1L}}{x_1} \\
0&\frac{m_{Q_1}}{x_2}
\end{pmatrix}
\nonumber\\
\overline V_L(k_1, s_1) V_R(k_2, s_2) &=&-\sqrt{k_1^+k_2^+}
\begin{pmatrix}
\frac{m_{Q_2}}{k_2^+}&0 \\
\frac{k_{1R}}{k_1^+}-\frac{k_{2R}}{k_2^+}& \frac{m_{Q_2}}{k_1^+}
\end{pmatrix}=-\sqrt{x_1x_2}
\begin{pmatrix}
\frac{m_{Q_2}}{x_2}&0 \\
\frac{k_{1R}}{x_1}-\frac{k_{2R}}{x_2}& \frac{m_{Q_2}}{x_1}
\end{pmatrix}
\nonumber\\
\overline V_R(k_1, s_1) V_L(k_2, s_2) &=& -\sqrt{k_1^+k_2^+}
\begin{pmatrix}
\frac{m_{Q_2}}{k_1^+}&\frac{k_{2L}}{k_2^+}-\frac{k_{1L}}{k_1^+} \\
0&\frac{m_{Q_2}}{k_2^+}
\end{pmatrix}=-\sqrt{x_1x_2}
\begin{pmatrix}
\frac{m_{Q_2}}{x_1}&\frac{k_{2L}}{x_2}-\frac{k_{1L}}{x_1} \\
0&\frac{m_{Q_2}}{x_2}
\end{pmatrix}
\eea
with $k_{1,2}^+=x_{1,2}P^+$. 
For eikonalized longitudinal momenta $k_1^+\approx k_2^+$ commensurate with our use of the wilson lines, (\ref{UVX1}) simplify
\bea
\label{UVX2}
\overline U_L(k_2, s_2) U_R(k_1, s_1) &\rightarrow&
\begin{pmatrix}
m_{Q_1}&0 \\
\Delta_R& m_{Q_1}
\end{pmatrix}=m_{Q_1}{\bf 1}+\Delta_R\sigma^-\nonumber\\
\overline U_R(k_2, s_2) U_L(k_1, s_1) &\rightarrow&
\begin{pmatrix}
m_{Q_1}&-\Delta_L\\
0& m_{Q_1}
\end{pmatrix}=m_{Q_1}{\bf 1}-\Delta_L\sigma^+\nonumber\\
\overline V_L(k_1, s_1) V_R(k_2, s_2)&\rightarrow&
\begin{pmatrix}
-m_{Q_2}&0\\
\Delta_R& -m_{Q_2}
\end{pmatrix}=-m_{Q_2}{\bf 1}+\Delta_R\sigma^- \nonumber\\
\overline V_R(k_1, s_1) V_L(k_2, s_2) &\rightarrow&
\begin{pmatrix}
-m_{Q_2}&-\Delta_L\\
0&- m_{Q_2}
\end{pmatrix}=-m_{Q_2}{\bf 1}-\Delta_L\sigma^+
\eea
with the momentum transfer $\Delta^\mu=k_1^\mu-k_2^\mu$, $\Delta_L=\Delta^1-i\Delta^2=\Delta_R^*$, and
$\sigma^\pm=\frac 12 (\sigma^1\pm i\sigma^2)$.
Inserting (\ref{UVX2}) into (\ref{TNL2}) yields  the local determinantal 2-body interaction potential (\ref{TNL2})
on the light front as

\bea
\label{27X}
&&V^{\eta^\prime}_{TH}\approx  - |\tilde\kappa_2| A_{2N} \frac 14 \big(1-\tau_1\cdot \tau_2\big)\,\nonumber\\
&&\times\bigg[ 4m_{Q_1}m_{Q_2}\,{\bf 1}_1{\bf 1}_2
-2\big(\sigma_{1\perp}\cdot i\nabla_\perp\,m_{Q_2}{\bf 1}_2-m_{Q_1}{\bf 1}_1\,\sigma_{2\perp}\cdot i\nabla_\perp\big) 
+\sigma_{1\perp}\cdot \sigma_{2\perp}\nabla^2_\perp 
\bigg]
\delta(P^+z^-)\delta(x_\perp)\nonumber\\
\eea
in the U(1) or $\eta^\prime$ channel, where only the leading $1/N_c$ contribution is shown. We have set
\bea
\label{27XY}
A_{2N}=\frac{2N_c-1}{2N_c(N_c^2-1)}\qquad\tilde\kappa_2=3!G_{\rm Hooft}\langle\bar s s\rangle<0\qquad G_{\rm Hooft}=
\frac{n_{I+\bar I}}2 (4\pi^2\rho^3)^3\prod_{f=u,d,s}\frac 1{m_f^*\rho}
\eea
\end{widetext}
with  $\sigma_{i\perp}$ the $\perp$-Pauli matrices  for particle $i=1,2$.
The flavor permutation $P^f_{12}$  is manifest in (\ref{TNL2}) as carried in (\ref{27XY})
$$\frac 14 \big(1-\tau_1\cdot \tau_2\big)=\frac 12(1-P^f_{12})$$ 
This is just the projector on the flavor singlet states. This is at the origin of the famed t${}^\prime$Hooft determinantal 
interaction, which helps solve the U(1) problem for the 3-flavor case. (Note that the interaction is repulsive in the un-projected 3-flavor U(1) channel).

Both spin contributions in (\ref{27X})) flip
the spin of the incoming quark pair from L-down to R-up in the instanton contribution, and vice-versa in the anti-instanton 
contribution. In the ultra-local approximation, we may trade $\nabla_\perp^2\rightarrow 1/\rho^2$,
hence  the instanton induced spin-spin interaction in the 2-flavor singlet channel,  

\bea
\label{THOOFTSS}
-  |\tilde\kappa_2| A_{2N} \frac 12 \big(1-\tau_1\cdot \tau_2\big)\,
S_{1\perp}\cdot S_{2\perp}\delta(P^+z^-)\delta(x_\perp)\nonumber\\
\eea

The nature of the instanton induced interactions in the other meson channels with different spin-flavor,
 is manifest in the individual contributions in the Fierzed form (\ref{TNL2}), with
{\it no contribution} to the vector and pseudo-vector channels. 
For instance, in the pion channel,
the light front interaction is

\bea
\label{THOOFTPION}
 |\tilde\kappa_2| A_{2N} \frac 14 \tau_1\cdot \tau_2\,
S_{1\perp}\cdot S_{2\perp}\delta(P^+z^-)\delta(x_\perp)
\eea
which is attractive in the isospin-triplet and spin-singlet state,  key for a massless pion. This point
has been addressed  in  the  literature many times before, including briefly  in this series~\cite{Shuryak:2021hng}
(note that in the latter the interaction was assumed local in the invariant distance $\xi_x$ for simlicity).
If  we were to treat this interaction perturbatively, it may appear that  we  should fit
the magnitude of the 4-fermion coupling constant to put the total pion mass to zero.
However, this is not correct. This effect is dominant  and leading in the pion channel, 
and should not be treated perturbatively, as we showed qualitatively in~\cite{Shuryak:2021hng}. For any large enough coupling
(and thus instanton density) it breaks chiral symmetry and (in the chiral limit)
produces  exactly massless Nambu-Goldstone modes, the pions. Yet to show that
quantitatively, we need to re-derive the essentials of chiral symmetry breaking 
 on the light front, a task we will discuss later in these series.

\newpage
\section{Various observables}~\label{sec_OBSERVABLES}

\subsection{Parton distribution functions}

The parton distributions functions  or PDFs count the parton content of a given hadronic state.
They are matrix elements of various operators sandwiched between pertinent light front
wavefunctions. For the hadronic states (\ref{WFPS}-\ref{WFVX}) limited to the lowest 2-particle
wave-function contributions with net helicity zero, the pseudo-scalar and vector PDFs are given by

\bea
\label{PDFX}
q_{P,V}(x)=\int {d^2 k_\perp \over (2\pi)^3} \big( |\psi^{P,V}_0|^2+ |\psi^{P,V}_1|^2+|\psi^{P,V}_{-1}|^2\big) \nonumber\\
\eea
with $q_{P,V}(\bar x)=q_{P,V}(x)$ for the anti-particles.
The PDFs normalize to  1 using  (\ref{NORMX}), which is  the charge sum rule for a
given hadron.  In the 2-particle approximation, the  momentum sum rule is automatically fulfilled
\bea
\label{SUM2}
\int_0^1 dx\, (xq_{P,V}(x)+\bar x q_{P,V}(\bar x))=1
\eea

\subsection{Distribution amplitudes}

The distribution amplitudes DAs are defined as matrix elements of certain nonlocal operators 
on the light cone, sandwiched between the vacuum and pertinent
hadronic states. They capture the longitudinal momentum, and transverse location of a parton
in the hadronic state.

DAs are widely used in the theory of hard exclusive reactions, such
as the hadronic form-factors in elastic scattering, and heavy meson semi-leptonic decays.  At high
momentum transfer, factorization allows a split of the scattering amplitude
 into   a ``hard block operator", sandwiched  between two  DAs. 
The moments of various DAs have been calculated on the lattice, for a reviews see \cite{Braun:2015lfa}.

The DAs are classified by the $twist$ (dimension minus spin) of the operator involved.
To give an example, since the early 1980$^\prime$s, most exclusive reactions involving the pions are based on the following three DAs
\begin{widetext}
\begin{eqnarray}
	\label{WF1}
	&&\int_{-\infty}^{+\infty}\frac{p^+dz^-}{2\pi}e^{ixp\cdot z}\left<0\left|\overline{d}_\beta(0)[0,z]u_\alpha(z)\right|\pi^+(p)\right>\nonumber\\
	&&=
	\bigg(+\frac{if_\pi}4 \gamma^5\bigg(\slashed{p}\,\varphi^A_{\pi^+}(x)
	-\chi_\pi \varphi_{\pi}^P(x)-i \chi_\pi
	\sigma_{\mu\nu}\frac{p^\mu n^{\nu}}{p\cdot n}  {\varphi_{\pi}^{\prime\,T}(x)\over {6}}\bigg)\bigg)_{\alpha\beta}
\end{eqnarray}
\end{widetext}
with $n^\nu$ a light-like vector in the z-direction.
Note that although these matrix elements are nonlocal,
the integral is just one-dimensional, taken along the light cone coordinate.
 The symbol $[x,y]$  is the shorthand notation for the gauge link between two points
 on the light front, and  $\sigma_{\mu\nu}=\frac 1{2i}[\gamma_\mu, \gamma_\nu]$. 
  In this example
 the first term contains momentum, while the other two do not. Therefore the axial $A$-DA is of leading twist, while the
 two others $P,T$-DA  are subleading (next twist) at large momentum $p\rightarrow \infty$.  
  
  The three functions $\varphi^i(x)$ have indices  $i=A,P,T$ standing for axial, pseudoscalar and tensor 
  gamma matrices in the operator.  They are all normalized to 1. Their explicit definition follows from (\ref{WF1}) by inversion

\begin{widetext}
\begin{eqnarray}
\label{PIDIS}
	&&\varphi^A_{\pi^+}(x)=
	\frac  1{if_\pi}\int_{-\infty} ^{+\infty} \frac{dz^-}{2\pi}e^{ixp\cdot z}\left<0\left|\overline{d}(0)\gamma^+\gamma_5[0,z]u(z)\right|\pi^+(p)\right>\nonumber\\
	&&\varphi^P_{\pi^+}(x)=
	\frac  {1}{f_\pi\chi_\pi}\int_{-\infty} ^{+\infty}  \frac{p^+dz^-}{2\pi}e^{ixp\cdot z}\left<0\left|\overline{d}(0)i\gamma_5[0,z]u(z)\right|\pi^+(p)\right>\nonumber\\
	&&{\varphi^{T\prime}_{\pi^+}(x)}=
	\frac  {-6}{f_\pi\chi_{\pi}}\frac {p^\mu n^\nu}{p\cdot n}\int _{-\infty} ^{+\infty} \frac{p^+dz^-}{2\pi}e^{ixp\cdot z}\left<0\left|\overline{d}(0)\sigma_{\mu\nu}\gamma_5[0,z]u(z)\right|\pi^+(p)\right>
\end{eqnarray}
\end{widetext}
 
\begin{figure}[htbp]
\begin{center}
\includegraphics[width=6cm]{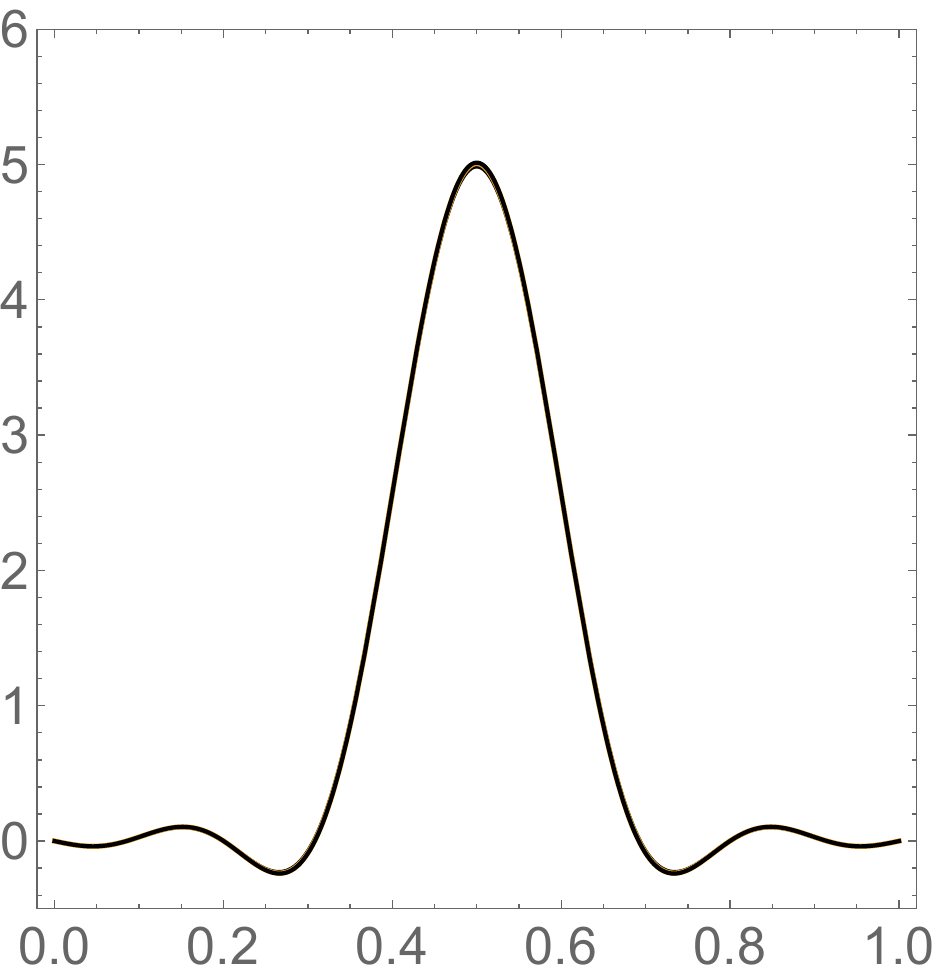}
\includegraphics[width=6cm]{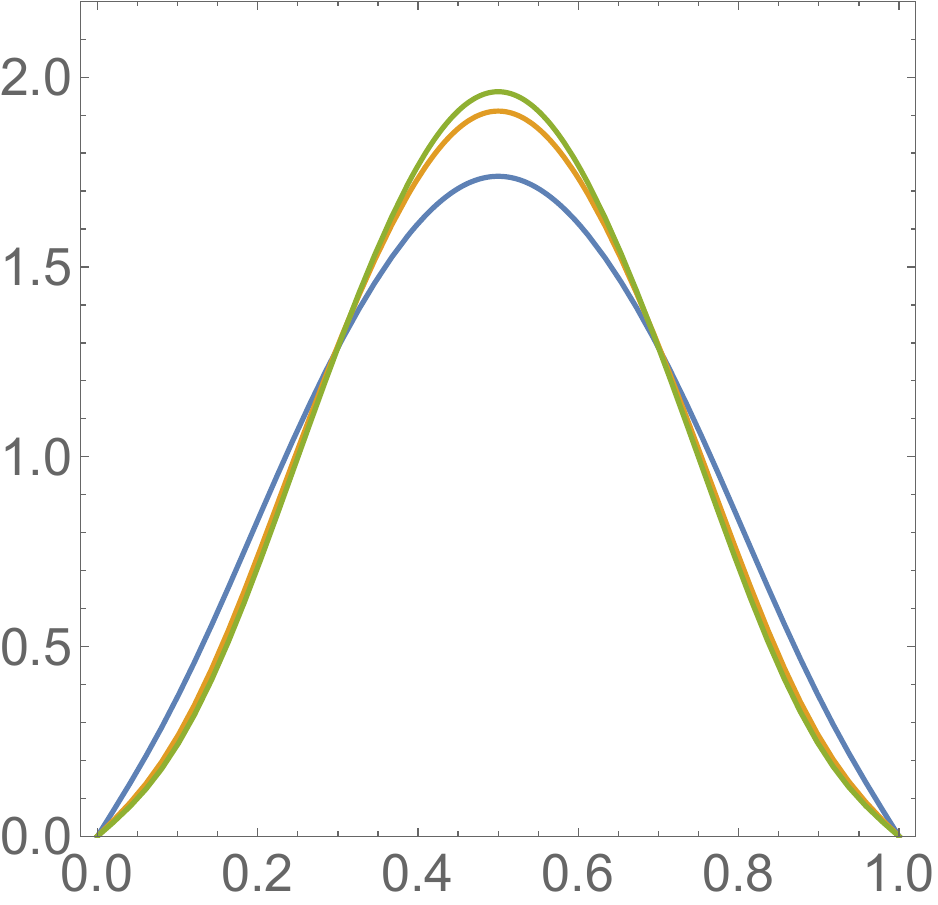}
\caption{ Distribution amplitudes for $\bar b b$ (upper) and a  ``generic" light $\bar q q$ meson (lower) as a function of $x$,
for the three lowest states $n=0,1,2$. For bottomium, the difference between the  three curves  is too small to be
visible. For the  light meson, the differences are visible.
With  increasing $n$, the DAs become narrower and higher at
$x=\frac 12$.}
\label{fig_3DAs}
\end{center}
\end{figure}

The constants DAs are normalized to 1. The axial pion DA is normalized
by the weak pion decay constant $f_\pi\approx 133\,{\rm MeV}$, and the pseudoscalar
and tensor pion DAs are normalized by $\chi_\pi$ which follows from the chiral algebra.
These chiral parameters are a measure of the pion wavefunction at the origin of
the {\it transverse plane} (zero transverse distance $b_\perp=0$). 
The DAs  are only a function of the longitudinal momentum $x$.
 \\
 \\
 {\bf Pion axial DA:}
 \\
To relate the pion axial DA  in (\ref{PIDIS}) to the pseudoscalar light front wavefunction
with helicity-0 in  (\ref{WFPS}), we note the identity

\bea
&&\tau^+_{ud}\bigg(\frac 14 \gamma^5\slashed{p}\bigg)_{\alpha\beta}=\nonumber\\
&&\frac 12 \frac 1{\sqrt{4x\bar x}}\big( u_{\alpha\uparrow}(1) \bar d_{\beta\downarrow} (2)- 
u_{\alpha\downarrow} (1) \bar d_{\beta\uparrow} (2)\big)
\eea 
in the light front limit. With this in mind, 
the leading twist-2  or axial distribution amplitude for $P=\pi^+$  (or any pseudoscalar) in (\ref{PIDIS}), 
matches the $m=0$  contribution in (\ref{WFPS}) 

\bea
\varphi^A_{P}(x)=\frac{2\sqrt{N_c}}{f_{P}}\int \frac{dk_\perp}{(2\pi)^3}\,\psi_0^{P}(x, k_\perp)
\eea
By the same reasoning, the corresponding contribution for the vectors,  
matches the $m=0$  contribution in (\ref{WFVX})

\bea
\varphi_V(x)=\frac{2\sqrt{N_c}}{f_V}\int \frac{dk_\perp}{(2\pi)^3}\,\psi_0^{V}(x, k_\perp)
\eea
\\
\\
 {\bf Pion tensor DA:}
 \\
To relate the tensor pion DA amplitude in (\ref{PIDIS}) to the pseudoscalar light
front wavefunction with  helicity-1 in (\ref{WFPS}), we note that the dominant tensor
matrix element on the light cone reads

\bea
&&\langle 0|\overline{d}(0)i\sigma^{+i}\gamma_5u(z^-, z_\perp)|\pi^+(p)\rangle\nonumber\\
&&=2p^+\frac{\partial}{\partial z_\perp^i}\psi^P_1(z^-, z_\perp)
\eea
with ($x=k^+/p^+$)

\bea
\psi_1^P(x, k_\perp)=
\int dz^-dz_\perp e^{i(k^+z^--k_\perp \cdot z_\perp)}\psi_1^P(z^-, z_\perp)\nonumber\\
\eea
 Note the relation  $\psi^P_{\pm 1}(x, k_\perp)=k_\perp^\pm \psi_1^P(x, k_\perp)$ 
with $k_\perp^\pm =k_x\pm ik_y$. 
A comparison with (\ref{PIDIS}) shows that the twist-3 pion distribution amplitude
in $\varphi_T(x, k_\perp)$ matches the $m=1$ contribution in (\ref{WFPS}) through

\bea
\frac{\partial}{\partial k_\perp^i}\varphi_T(x, k_\perp)=\frac{6}{f_\pi\chi_\pi}k_\perp^i\psi^P_1(x, k_\perp)
\eea
 \\
 \\
 {\bf Bottomium DA:}
 \\
The generic LFWFs depend  on all three  variables $x, b_\perp$.  
The DAs are LFWF's  at the origin in the coordinate representation, or the overall integral

\begin{equation}
\label{FOURB}
	\Psi_{nm}(b_\perp=0, x)=\int {d^2 k_\perp \over (2\pi)^2} \Psi_{nm}(k_\perp,x)
\end{equation} 
in the momentum representation.
They are normalized so that the  integral over $x$ is one. 
In Fig.~\ref{fig_3DAs} we show the DAs for $n=0,1,2$ LFWFs, for $\bar b b$ and generic $\bar q q$
mesons. 

Note first, that the heavier  the meson, the slower the quark motion, with the DAs
concentrated near $x=\frac 12$. Note also, that  the bottomonia states with
different orbital momenta  $m$, have practically the same DAs. Note finally, that the oscillations
in the DAs for  bottomonia, are an artifact of the  small basis set of functions in $x$, we used.
Before normalization, the lowest  bottomium DA reads
\bea
&& DA(n=0,m=0) \sim \big(2.19 sin(\pi x) \nonumber\\
&&- 
    1.79 sin(3\pi x)+ 
    1.16 sin(5\pi x)- 
    0.44 sin(7 \pi x) \big). \nonumber 
\eea
Note that the signs of the harmonics alternate, with a net effect being a suppression of  the DA near the edges $x\rightarrow 0,1$.
\\
\\
\noindent{\bf Decay constants:}
\\
These matrix elements are best interpreted in the chiral basis.
For the pseudoscalar pion,  the leading twist-2 DA  $\varphi_\pi(x)$ is {\em chirally-diagonal}, but the 
subleading  twist-3 are {\em chirally non-diagonal}. Fortunately, a matrix related the chiral basis
to the spin basis has been already defined in the previous section. 

The weak pion decay constant follows 

\begin{widetext}
\begin{equation}
	\label{WF11}
	\left<0\left|\overline{d}(0)\gamma^\mu{(1-\gamma^5)}u(0)\right|\pi^+(p)\right>=
	-{\rm Tr}\bigg(\gamma^\mu{(1-\gamma^5)}\bigg(\frac{i}4\gamma^5\slashed{p}\bigg)\bigg)\,
	2\int \frac{dk_\perp}{(2\pi)^3} \frac{dx}{\sqrt{N_c}}\,\psi_0^{P}(x, k_\perp) \equiv 
	if_\pi\,p^\mu
\end{equation}
\end{widetext}
with the trace referring to color-spin. The  generic pseudo-scalar weak decay constant is then

\bea
f_{P}=2\sqrt{N_c}\int \frac{dk_\perp}{(2\pi)^3} dx \,\psi_0^{P}(x, k_\perp)
\eea

Similar arguments applied to the weak decay constant of the $V=\rho^+$ meson defined as

\bea
\left<0\left|\overline{d}(0)\gamma^\mu u(0)\right|\rho^+(p)\right>=\epsilon^\mu (p)m_\rho f_\rho
\eea
yield the generic vector decay constant

\bea
f_{V}=2\sqrt{N_c}\int \frac{dk_\perp}{(2\pi)^3} dx \,\psi_0^{V}(x, k_\perp)
\eea
\\
\\
{\bf Chiral   constant:}

Isospin symmetry and charge conjugation force
$\varphi_\pi(x)=\varphi_\pi(\overline x)$.
As pointed out initially in  \cite{Geshkenbein:1982zs}, there are two twist-3 and {\em chirally non-diagonal} independent  DA 
$\varphi^P_\pi(x)$ and $\varphi_\pi^T(x)$, characterizing the pseudoscalar and tensor strength in the pion respectively.  The
latters are tied by the current identity

\bea
&&\partial^\nu\big(\overline{d}(0)\sigma_{\mu\nu}\gamma_5u(z)\big)\nonumber\\
&&=-\partial_\mu\big(\overline{d}(0)i\gamma_5 u(z)\big)+m\,\overline{d}(0)\gamma_\mu\gamma_5u(z)
\eea
and share the same couplings.
The value of the  dimensionful coupling constant $\chi_\pi$ can be fixed by the divergence of the axial-vector  current
and the PCAC relation

\begin{widetext}
\bea
 \label{eqn_div_of_axial}
	&&(m_u+m_d) \left<0\left|\overline{d}(0)i\gamma^5u(0)\right|\pi^+(p)\right>=\nonumber\\
	&&-(m_u+m_d)\,{\rm Tr}\bigg(i\gamma^5\bigg(\frac{if_\pi}4\gamma^5 \chi_\pi\bigg)\bigg)\,\int_0^1dx\,\varphi_{\pi}^P(x)
	=(m_u+m_d)\,f_\pi\chi_\pi
\eea
\end{widetext}
with $\varphi_{\pi}^P(x)$  normalized to 1.  Using the Gell-Mann-Oakes-Renner  relation
\begin{equation}
	f_\pi^2m_\pi^2=-2(m_u+m_d)\left<\overline{q}q\right>
\end{equation} 
with $|\left<\overline q  q\right>|\approx (240\,{\rm MeV})^3$, which yields
\begin{equation}
	\label{CHI}
	\chi_\pi=
	\frac{m_\pi^2}{(m_u+m_d)}
\end{equation}

\subsection{Form-factors}

Elastic scattering is the simplest exclusive process, and the corresponding mesonic
formfactors have been studied extensively theoretically and experimentally, for 
about five decades. Here we do not have space for a review of their history,
let us just say that early asymptotic predictions at large $Q^2$, based on one-gluon exchange,
are not yet met, neither in experiment or on the lattice. 

(The ``semi-hard" domain  $Q^2\sim few\, GeV^2$
is dominated by nonperturbative effects. 
In particular, in
our previous paper \cite{Shuryak:2020ktq},  we have calculated the instanton-induced contributions
to the hard block, and showed that they are comparable to the
perturbative amplitudes. When combined,  they reproduce the  experimental/lattice data, provided 
that the diluteness parameter is not small $\pi^2 n_{I+\bar I} \rho^4 =O(1)$.)

On the light front, the form factors follow from the  Drell-Yan-West construction using the good current
$J^+=J^0+J^3$~\cite{Drell:1969km,West:1970av}. The analogue
of the Breit-frame with fixed energy in the rest frame, is the Drell-Yan frame in the light front frame, with fixed 
longitudinal momentum $P^+=P^{\prime+}$, for $P^\prime=P+q$ with space-like squared momentum
transfer $q^2=-Q^2=-q_\perp^2$.  The key feature of this choice of current and frame, is that the
vacuum production and annihilation diagrams are suppressed, and parton number is conserved in-out
in the form-factor viewed as a process $\gamma^*+P\rightarrow P^\prime$. 

For instance, 
the Drell-Yan-West form-factor  for charged pseudo-scalars is 
helicity preserving with~\cite{Drell:1969km,West:1970av, Lepage:1980fj}

\bea
&&F_P(Q^2)=\langle P^\prime, 0,0|\frac{J^+(0)}{2P^+}|P, 0,0\rangle \nonumber\\
&&=\int_0^1 dx\int \frac{dk_\perp}{(2\pi)^3} \psi^{P*}_0(x, k_\perp+\bar x q_\perp)\psi_0^P(x, k_\perp) \nonumber\\
\eea
in the 2-particle approximation (low resolution).
Similar expressions for the charged vectors can be derived. The thorough analyses of these
form factors will be given in a sequel.

\section{Conclusions}~\label{sec_CON}

The chief aim of this series of papers is to $derive$ the light-front  Hamiltonian $H_{LF}$ and corresponding wave functions
from first principles, using the theoretical and empirical information we have at the moment. The important distinctions between our approach and other versions of $H_{LF}$ 
in the literature are among others: (i) our $H_{LF}$ is not a model, meaning  we do not invent its  form, but derive
it  using certain approximations; (ii) therefore we do not fit any of the parameters to the experimental data, rather
we use the standard values for the quark masses, the string tension $\sigma_T$, the perturbative coupling $\alpha_s$ and  the
instanton ensemble parameters;
(iii) we do not consider just one or two lowest states in each quark channel, but
look for as many states as feasible, by relating the results to Regge behavior;
(iv) central to our derivation,  is the QCD vacuum as we know it, at low resolution.

In a wider perspective, these works continue to bridge light front  physics with the development in Euclidean space-time. The latter -- 
lattice QCD simulations and semiclassical ensembles of instantons --  
have elucidated a rich vacuum structure dominated by inhomogeneous 
and topologically nontrivial gauge configurations. These configurations  explain why chiral symmetry 
is spontaneously broken, and account for the emergence of mass through running quark effective masses.

A massless left-handed quark tunneling through an instanton
emerges as a right-handed massless quark as a topological zero mode, a remarkable manifestation of the of axial anomaly. This phenomenon is the essence of the dynamical
breaking of chiral symmetry, which yields a running constituent mass. The collectivization of these zero modes, is well undertstood from detailed numerical studies
of instanton ensembles carried in the  1990's, and
produces an  octet of massless Nambu-Goldstone modes. The QCD vacuum possesses  the ``Zero Mode Zone" with no gap near zero Dirac eigenvalue.
In a way, we may say that the QCD vacuum is $^\prime metallic^\prime$, with the scalar and 
vector mesons as weakly correlated  $^\prime excitons^\prime$. 

However, chiral symmetry breaking is not the only instanton-induced effect.
Apart of well-isolated instantons producing near-zero Dirac modes, there are also more numerous
fluctuations which can be described as instanton-antiinstanton molecules.
In the first paper of the series \cite{Shuryak:2021fsu},  we showed  that by including them
in Wilson line correlators,  we  can account for a significant part of the central inter-quark potential,
as well as the spin-dependent potentials in heavy quarkonia.

The light front  observables at large normalization scale $\mu^2$ paint a picture of hadrons
 containing multiple gluons, and a  rich quark-antiquark sea. Yet at 
   small resolution, of the order of $\mu \sim 1/\rho\approx 600\, MeV$, we expect
this picture to morph into the  spectroscopic quark model, dominated by the minimal (2 for mesons, 3 for baryons) configurations.   
   This is  especially obvious for heavy quarkonia, which is the focus in this paper.
 
  ``Potentials" independent and dependent on spin variables,  are defined via
  certain {\em nonlocal} observables, such as the well known Wilson lines.
  We know how to evaluate  them 
   in the Euclidean version of the QCD vacuum, on the lattice or when it can be regarded
as a correlated ensemble of certain topological solitons, such as instantons and  anti-instantons.  

Bridging Euclidean vacuum with the light  front,  is one of the major challenge for theory today.
In  the second paper \cite{Shuryak:2021hng}, we outlined   a method to do so, by
performing calculations in Euclidean space-time,  and then analytically
continuing  {\em the results} (not the field configurations themselves).
Wilson loop correlators are mapped on the light front, in terms of a 
 slope angle $\theta$ in Euclidean signature, that is continued to  $-i\chi$  (the rapidity) in Minkowski
signature. This  proposal was made long ago, and tested both at the perturbative~\cite{Meggiolaro:1997mw}, 
and  non-perturbative~\cite{Shuryak:2000df,Giordano:2009su}  levels.

 In a way, this proposal is different, although similar in spirit, to the large momentum effective 
approach~\cite{Ji:2013dva}, where Euclidean equal-time correlators are made to asymptote their light cone counterparts. The large momentum limit in this case,
is analogous to the large rapidity $\theta\rightarrow -i\chi$ continuation in our case. 

Finally, in this paper we explore the fortunate fact that
on the light front, all hadrons -- from bottomonia to the pions -- can be approached
from the same LF Hamiltonian. In the first approximation, it is just a transverse oscillator
and longitudinal harmonic functions. In the next approximation, the non-factorizable part $\tilde V$
is included.  For the lowest states, its influence is not too drastic.

The next approximation brings in the Coulomb, perturbative and instanton-induced
mixing, in spin and angular momenta. For heavy quarkonia these effects  are
suppressed by large quark masses. For light quark states, these effects 
are also not  large, except for the ground states. The reason  is that
these mixing effects are short-range.
Narrow Wilson loops are mostly sensitive to the topologically active instanton and anti-instanton 
gauge  configurations and chiral symmetry breaking. Wide Wilson loops are mostly sensitive to  the flux disordering gauge vortices and confinement.
Contrary to common lore,  the QCD vacuum is dominant on the light front too, and is central for the emergence of mass, confinement and spin mixing.

The idea to derive $all$ LF observables -- DAs, PDFs, GPDs, Form-factors -- from a common model Hamiltonian, 
was put forth by  Brodsky and his collaborators~\cite{Brodsky:2014yha}, and by Vary  and his collaborators~\cite{Li:2015zda,Mondal:2021wfq}, 
with the intent  to combine results of various experiments into a common framework.  Their Hamiltonians
were largely guessed, using insights  from holographic QCD models. 

Our approach is different. It is  based on a derivation of a light front Hamiltonian, from  well established
features of the perturbative and nonperturbative QCD vacuum fields.  We also 
used a  wider set of states and quark masses, and showed that the excited states
are much closer to the generic Regge trajectories. Indeed, for the excited states, the confining string becomes the only 
relevant physical effect, as explained in our introductory remarks in~\cite{Shuryak:2021fsu}.

Looking at all works on $H_{LF}$, 
we  believe that this is just  the  beginning
of a successful modeling of hadrons on the light front,
that factors in the wealth of information from the QCD vacuum. 
While our analyses deal solely with mesons, 
their  generalization to baryons
is relatively straightforward. Also, one can analyze multi-quark wavefunctions of mesons (tetraquarks)
and baryons (pentaquarks), the anti-quark sea etc. 

In many ways, the theoretical construction and tools presented and used in these
serial analyses, provide a common  framework for both hadronic and nuclear physics.

\vskip 1cm
{\bf Acknowledgements}

This work is supported by the Office of Science, U.S. Department of Energy under Contract No. DE-FG-88ER40388.

\appendix

\section{Functional basis using a 2d oscillator Hamiltonian}~\label{2d_basis}

Following our analysis in~\cite{Shuryak:2021hng}, we use the eigenfunction basis of the ``2d oscillator" imbedded in part of the
Hamiltonian $H_0$ in (\ref{H0X}). We start by recalling its generic properties,
and then use it either in the momentum or  coordinate representation, whichever is  more
convenient.

The generic Hamiltonian is 
\be H_{osc}=\vec p^2 {1\over 2 \mu}+ \vec \rho^2 {\mu \omega^2\over 2} \ee  
where $\vec\rho$ is a 2d coordinate. One way to generate all wave functions is to
use two 1d oscillator notations, but a  more convenient one is to use polar coordinates.

 so the Hamiltonian matrix consists of separate sectors of $N=3\times 4$ size.

The basic LF Hamiltonian includes a diagonal $H_0$ and a nonfactorizable part

\be 
\tilde V=(M^2 + \vec p_\perp^2)\bigg({1 \over x(1 - x)} - 4\bigg)
\ee
which can be calculated either in momentum or coordinate representations.

Note that  the wave functions we use here,  are all normalized using
$$\int_0^1 dx\int  d^2\rho_\perp \psi_{nm}^2(x, \rho_\perp)=1\,,$$ which is natural in coordinate space.
When used in momentum space, pertinent  powers of  $1/(2\pi)^3$ will be added 
whenever  needed.

The wave functions which are independent of the azimuthal  angle (angular momentum zero) will be used
mostly inthe  momentum representation,  with $\vec \rho$ as the transverse momentum. 
The orthonormal wave functions are

\bea
\label{eqn_psi_n0}
&&\{\psi_{n0}) \} =  e^{- \beta^2 \rho^2/2}   \sqrt{1\over \pi}\beta\nonumber\\
&& \times\{1,(1 - \beta^2 \rho^2), (1 - 
     2 \beta^2 \rho^2 + \beta^4 \rho^4/2),... \} \nonumber\\
\eea
with the $\beta$-parameter given in terms of the Hamiltonian parameters $\beta=  (4a/b/ \sigma_T)^{1/4} $.  
     
The harmonic set of orthonormal  functions for longitudinal momentum fraction $x$, 
the set of functions $\chi_l(x)$, are labeled by $odd$ integer  $l=1,3,5...$,  

\bea
&&\chi_l(x)= \sqrt{2}\nonumber\\
&&\times\{   Sin(\pi x), \,
  Sin(3 \pi x), 
    Sin(5 \pi x), 
   Sin(7 \pi x),... \} \nonumber\\
   \eea
The products of these two sets,  define the set of states we used fin our (angle-independent) calculations,
in the momentum representation.   

 The part of the Hamiltonian matrix used is thus limited by three maximal
 values of indices $n,m,l$, so the total size of of the matrix is $N\times N$,
 with $N=n_{max} m_{max} l_{max} $. Obviously, the calculations significantly slow down
 with increase $N$, and so for this exploratory paper we use 
a  rather modest value of $N=3\times 3 \times 4$. Furthermore,  before we
 account for mixing of states with different orbital momenta, $m$ is conserved,

\section{From bottomonium to generic light mesons, on the light front}~\label{bottom_LF}

In our study in~\cite{Shuryak:2021hng},  we focused on the light quark systems on the light front.
Here we start with  the other extreme with the example of  $\bar b b$ states, for which we use for
the  parameter $b= (2*4.8 \, GeV)^2; m_b = 4.8\, GeV$. 
In this section, we start  with the states 
 with zero angular momentum or $m=0$.  Our $M^2(a)$ curves lead to the selection of  mass minima at about  $a\approx 25$. 
With this in mind, we find  the following twelve eigenvalues for the squared mass

\begin{widetext}
 $$M^2 \approx \{360., 341., 328., 169., 161., 153., 127., 
121., 114., 113., 107., 101.\} \, (GeV^2) $$
The  LFWF of the ground state is  approximately factorized into

\be 
\Psi_{00}=e^{-1.302 p_\perp^2} \big(-0.915 Sin(\pi x) + 0.749 
      Sin(3 \pi x) - 
   0.485 Sin(5 \pi x)  + 0.183 Sin(7 \pi x) \big)\ee
   \end{widetext}
   where we have omitted all terms with coefficients smaller than $0.01$.
 Note that the ground state   is then the product of just a Gaussian in transverse momentum, times
 certain functions of $x$.  However, the  next eigenstates are not that simple. 
The LFWFs for the next two states  with $n=1,2$ (and still independent on $\phi$ or for $m=0$) are

\begin{widetext}
\bea
 \Psi_{10}&=&e^{-1.302\rho^2} \big[(0.915 - 2.410 \rho^2 + 
      0.025 \rho^4) Sin(\pi x) + (-0.749 + 
      1.973 \rho^2 - 0.018 \rho^4) Sin(3 \pi x) \nonumber \\ &+& 
  ( 0.485  - 1.279 \rho^2) Sin(5 \pi x) + 
  ( - 0.183  + 
   0.484\rho^2  - 
   0.0025 \rho^4) Sin(7 \pi x)\big]
   \eea
   
\bea
 \Psi_{20}&=&e^{-1.302\rho^2} \big[(0.905 - 4.726\rho^2 + 
      3.085 \rho^4) Sin(\pi x) + (-0.741 + 
      3.876 \rho^2 - 2.533\rho^4) Sin(3 \pi x)  \nonumber \\ &+& 
   (0.480 - 2.521 \rho^2  + 
   1.651 \rho^4) Sin(5 \pi x) - (0.182  + 
   0.957 \rho^2  - 
   0.628 \rho^4) Sin(7 \pi x) \big] 
   \eea
Here $\rho^2=\vec p_\perp^2 \, (GeV^2)$ and all coefficients are also in GeV units
with appropriate powers. The 
$\Psi_{n0}$ functions are 
 eigenstates of $H$, and should not  be confused with the basis set  $\psi_{nml}$ introduced above.   
\end{widetext}

Their integrals of $\Psi_{nm}$ over $p_\perp$ give the DAs  discussed in section~\ref{sec_OBSERVABLES}.
Their $x$-dependence are very similar. Their
$p_\perp$-dependence is shown in Fig.\ref{fig_pt_LFWFs}. 
In contrast to the $x$-dependence, the $p_T$-dependences are  very different, as each 
curve reflects on the  proper number  of $n$ zeros.

The same construction for generic light quarks  with $m_q=0.35\, GeV$, was discussed in~\cite{Shuryak:2021hng}.
Here, we slightly modify the setting by selecting the variational minima at $a=4$. 
The $p_\perp$ dependence of the first three states is  shown in~Fig.\ref{fig_pt_LFWFs}.

For reference, the squared masses of the 12 lowest eigenvalues are 
 
\ba 
M^2=26.20, 23.70, 21.92, 15.77, 13.87, 12.28, \nonumber \\
 8.79,7.21, 5.77, 4.63, 3.43, 2.23 \, (GeV^2)  \nonumber \ea
and the lowest wave function is 

\begin{widetext}
\bea
\Psi_{00}=&&
e^{-7.14286 \rho^2} ((-3.12914 + 1.86018 \rho^2) - 
      0.72094 \rho^4) sin(\pi x) \nonumber\\&&+ (0.16657 + 0.636644 \rho^2 - 
      0.490831 \rho^4) Sin[3 \pi  x] \nonumber\\&&+ 0.0267555 Sin[5\pi  x] + 
   0.111208 \rho^2 Sin[5\pi  x] - 
   0.152454 \rho^4 Sin[5\pi  x] \nonumber\\&&+ 0.00919945 Sin[7 \pi  x] + 
   0.0402368 \rho^2 Sin[7 \pi  x] - 
   0.0577019 \rho^4 Sin[7\pi  x])
   \eea
\begin{figure}[h]
\center
\begin{tabular}{ll}
\includegraphics[width=7.5cm]{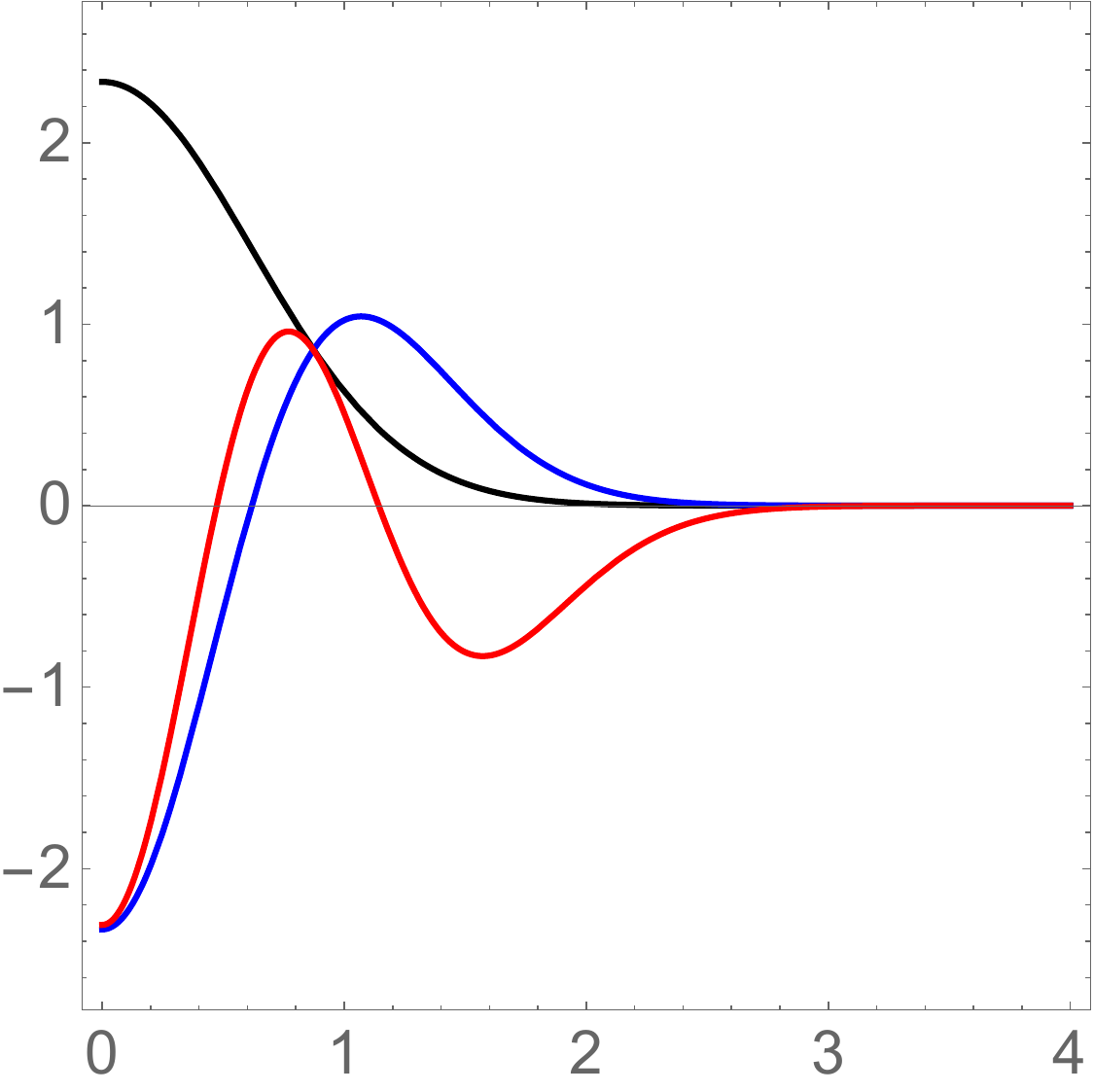} 
&
\includegraphics[width=7.5cm]{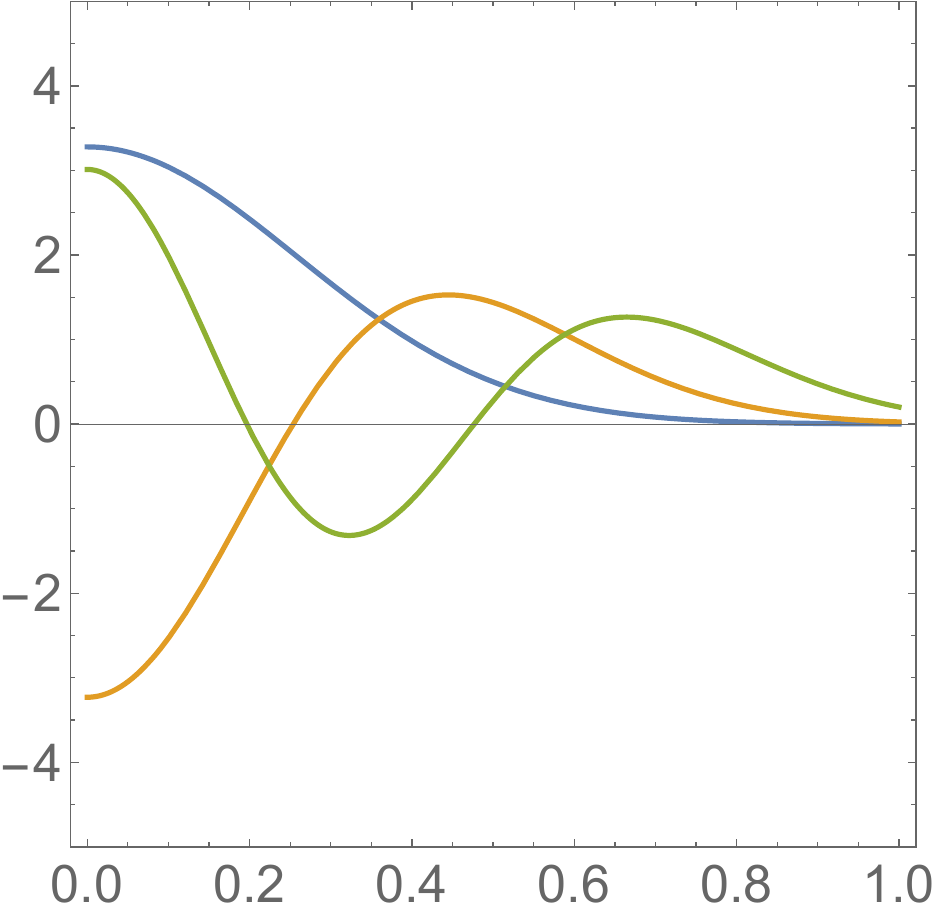}
\end{tabular}
\caption{Three lowest LFWFs
  with $n=0,1,2,m=0$  as a function of $p_\perp\, (GeV)$ at $x=\frac 12$, for bottomonium (left) and a typical light meson (right). 
  The number of zeros are commensurate with $n$.}
\label{fig_pt_LFWFs}
\end{figure}
\end{widetext}

\section{Wave functions with non-zero angular momentum $L_z=m$}~\label{wf_m}

The functions with nonzero 
orbital momentum $m$, are generated by the corresponding right (plus) creation operators 

\bea
 a^+_R&=& {1\over 2} e^{i \phi}\big(\beta \rho - {1 \over \beta} {\partial \over \partial \rho} 
-{i \over \beta \rho } {\partial \over \partial \phi} \big) \nonumber \\
a^+_L&=&{1\over 2} e^{-i \phi}\big(\beta \rho - {1 \over \beta} {\partial \over \partial \rho} 
+{i \over \beta \rho } {\partial \over \partial \phi} \big)
\eea  
and the needed extra factors for proper wave function normalization $1/\sqrt{n_R! n_L! } $
depending on the numbers of right- and left-rotating ``quanta".  

In particular, we use the following  orthonormal sets of functions $\psi_{nm}$ depending on  $\rho$ and azimuthal angle 
$\phi$, with principle quantum number $n=0,1,2...$
and angular momentum $m=0,1,2...$

\begin{widetext}
 \ba \label{eqn_psi_n2}
 \{\psi_{01} \}  &=& e^{- \beta^2 \rho/2+i\phi}   \sqrt{1\over \pi} \beta^2 \rho 
 \times   \{ 1,  (-2+\beta^2\rho^2)/\sqrt{2}, (6-6\beta^2\rho^2+\beta^4\rho^4)/2\sqrt{3} ,...\} \ea
  \ba \label{eqn_psi_n2}
  \{\psi_{0 2} \} &=&  e^{-\beta^2 \rho^2/2+2i\phi}   \sqrt{1\over 2\pi}\beta^3 \rho^2\times 
  \{1 ,(3-\beta^2\rho^2) /\sqrt{3},(12-8\beta^2 \rho^2+\beta^4\rho^4)
   {\sqrt{2}\over 4 \sqrt{3}} , ...  \} 
   \eea
   \end{widetext}
   
   When matrix elements of some potentials are evaluated, it is more natural to switch to the coordinate representation. One way to do it is to rederive
an oscillatory basis in which $\rho^2=\vec r_\perp^2$ and $\vec p^2$
is interpreted as a Laplacian containing angular (centrifugal) term.
In this case, the  parameter $\beta$ is inverted.
Simpler is to  go to coordinate representation by 2d Fourier transform. While
doing so, it is convenient to return to Cartesian
   coordinates, e.g. $\rho e^{\pm i\phi} \rightarrow  p_x\pm i p_y$, which after double Fourier
   transform produces factors $x \pm iy$. With slight abuse of notation, we write the latter combination as $r e^{\pm i \phi}$, although the angles $\phi$ in momentum and coordinate
   representations do not have the same meaning.
  Also note that in coordinate representation one may better use  
 the $inverted$ scale parameter
$$ \beta_p \rightarrow \beta_r= {1 \over\beta_p}= \bigg({4a \over b \sigma_T}\bigg)^{-\frac 14}, $$
Recall that $a$ is to be determined from the mass minimization, $b=M_{mes}^2\approx(2m_Q)^2 $,  and the string tension is  standard
 $\sigma_T = (0.4 \, GeV)^2$. Note also that, as expected, the wave functions at small distances
 are $\sim \rho^m$.
     
Now we return  to the momentum representation, and diagonalize $H_0+\tilde V$ for $m=1,2$.
For the bottomonium parameters the lists of the six lowest squared masses are 
\ba M^2_{0,\pm1}&=&\{ 131.1, 
124.4, 117.8, 116.6, 110.4, 104.2\}  \nonumber \\
 M^2_{0,\pm2}&=&\{134.4, 127.7, 121.1, 119.7, 113.5, 107.3 \} \nonumber
\ea

\begin{widetext}
\bea
 \label{eqn_bb_L12}
\Psi_{01}&= &
e^{-1.30 \rho^2}\rho e^{\pm i\phi} \big[(1.48- 0.015 \rho^2 + 
     ) sin(\pi x) \\
     &+& (-1.21 + 
      0.011 \rho^2 ) sin(3 \pi x) + 
   0.787 sin(5 \pi x)  - 0.298 sin(7 \pi x) \big], \nonumber \\
    \Psi_{02}&=&   e^{-1.30 \rho^2} \rho^2 e^{\pm 2 i  \phi}\big[ 1.67  
     sin(\pi x)  \\
     &-& 1.37 sin(3 \pi x) + 0.89 sin(5 \pi x) 
 - 0.34 sin(7 \pi x) \big]
   \eea
   \end{widetext}

\section{Mixing matrix elements}~\label{mixing_me}

In  (\ref{eqn_mixing}) we defined the  $3\times 3$ mixing matrix between 
states with  different azimuthal quantum numbers $m=0,1,2$, for a meson with fixed helicity $\Lambda=1$. For simplicity,
we did not use the states  $\Psi_{nm}(x,\vec k_\perp)$ determined in the previous section,
but rather the basic and simple oscillatory states $\psi_{nm}(k_\perp)$. These states carry the azimuthal 
dependence through $e^{i m\phi}$,  and the  $x$-dependence  through $\sqrt{2} sin(n\pi x)$,  which are standard and not 
explicitly shown.

In this simplified basis, the Coulomb interaction 
$$2MV_C= 2M_{\rm mes}\bigg(-{4\alpha_S \over 3 \rho}\bigg) $$
is diagonal
\bea
C^{mm} = \int_0^\infty |\psi_{0m}|^2\, V_C(\rho) 2\pi \rho d\rho
\eea
For  $m=\{0,1,2\}$, the entries are explicitly
$$ C^{mm}= M_{\rm mes} \sqrt{\pi} \alpha_S \beta \{-8/3, -4/3, -1\}$$ 
Note that here, we use the coordinate representation and the oscillator parameter $\beta$,
the inverse of the oscillator parameter in the momentum representation.

The perturbative spin-spin interactions is $ \vec S_{1} \vec S_{2}=\frac 14$,  for the states with total spin $S=1$, so
that $S_{1\perp}S_{2\perp}=\frac 23 \frac 14$.
Its associated transverse Coulomb potential $\nabla_\perp^2 /\rho$ is regulated at short distances, through
\bea 
V_{pert}^{SS} &=&\frac 23 \frac 14 \bigg({2M_{\rm mes} \over m_q^2}\bigg)
\bigg(-{4 \alpha_S\over 3}\bigg)\nabla_\perp^2 {1 \over \sqrt{\rho^2 + \epsilon^2}}
  \nonumber \\ &=&
  -\frac{4 M_{\rm mes} \alpha_S}{9m_q^2} {-2 \epsilon^2 + \rho^2 \over  (\epsilon^2 
+ \rho^2)^{5/2}} \eea
In the limit  $\epsilon \rightarrow 0$, it reduces to
\ba SS_{pert}^{mm}={2M_{\rm mes} \sqrt{\pi} \alpha_S \beta^3 \over m_q^2} \{4/9,-2/9,-1/18 \} \nonumber  \ea
for $m=\{0,1,2\} $

The induced instanton  spin-spin, spin-orbit and tensor forces cannot be carried analytically. We evaluated them numerically,
using the potentials shown in~Figs.\ref{fig_Vc_VT},\ref{fig_spin_orbit_inst}, as explained in the text.
The numerics are carried for light quarks with mass $m_q=0.35\, GeV$, and $M_{mes}=2m_q$. The matrix elements are
integrals of these potentials times the pertinent wave functions $\psi_{0m},m=0,1,2$ in coordinate space. 
The  results are given in  the second line of (\ref{eqn_mixing}).

\section{Spin, helicity and chirality spinors}~\label{sec_spinors}

The light-front wave functions in~\cite{Ji:2002xn} are built in terms of the spin and angular momentum
projected along the $z$-direction,  which is the  $hadron$ momentum  $\vec P$ direction. In the case of mesons,
there are two sets of spin variables $S_{Q,\bar Q}=S_{1,2}$
 and a single orbital momentum $L$.

Let the direction of  the quark momentum be described by standard
polar angles $\theta,\phi$, with $p_\perp=P sin(\theta)$ etc. In this case the spin up and down basis (with standard Dirac
matrices)  is

\begin{equation}
| spin \uparrow \rangle =\sqrt{E + m \over 2 m} \times	\begin{bmatrix}
		1 \\
		 0\\
		  {p \over E + m} cos(\theta) \\
		 {p \over E + m} sin(\theta)
			e^{i\phi}
	\end{bmatrix}
\end{equation} 
\begin{equation}
	| spin \downarrow \rangle =\sqrt{E + m \over 2 m} \times	\begin{bmatrix}
		0 \\
		1\\
		{p \over E + m} sin(\theta)
		e^{-i\phi} \\
		{p \over E + m} cos(\theta) 		
\end{bmatrix}
\end{equation} 

The helicity $\lambda=\vec S \cdot \vec k$ defines a different basis,  because 
the  spin projection is defined not along the in-coming $z$-axis,
but along the quark momentum.  Of course, quarks in the hadron 
 have nonzero transverse components to it, $ | p_\perp |=p sin(\theta)$. 
 The nonrelativistic 2-component spinors with $\lambda=\pm 1$ are obtained by rotation
\begin{eqnarray} h_+&=&\big( cos(\theta/2), e^{i\phi} sin (\theta/2) \big)  \\
 h_-&=&\big( -sin(\theta/2), e^{i\phi} cos(\theta/2) \big)   \nonumber
\end{eqnarray}
and, after a boost, the corresponding Dirac spinors are
\begin{equation}
	| h + \rangle =\sqrt{E + m \over 2 m} \times	\begin{bmatrix}
		cos(\theta/2) \\
		sin(\theta)
		e^{i\phi}\\
		{p \over E + m} cos(\theta/2) \\
		{p \over E + m} sin(\theta/2)
		e^{i\phi}
\end{bmatrix}
\end{equation} 
\begin{equation}
| h - \rangle =\sqrt{E + m \over 2 m} \times	\begin{bmatrix}
	-sin(\theta/2) \\
	cos(\theta/2) e^{-i\phi}\\
	{p \over E + m} sin(\theta/2)
\\
-{p \over E + m} cos(\theta/2) e^{-i\phi} 	
\end{bmatrix}
\end{equation} 

We will also use the chiral basis, which is obtained from
the helicity  basis  by taking the ultrarelativistic limit
($m\rightarrow 0, p/(E+m)\rightarrow 1$) inside the spinor 

\begin{equation}
	| c + \rangle =\sqrt{E + m \over 2 m} \times	\begin{bmatrix}
		cos(\theta/2) \\
	sin(\theta)
	e^{i\phi}\\
 cos(\theta/2) \\
 sin(\theta/2)
	e^{i\phi}
\end{bmatrix}
\end{equation} 
\begin{equation}
| c - \rangle =\sqrt{E + m \over 2 m} \times	\begin{bmatrix}
-sin(\theta/2) \\
cos(\theta/2) e^{-i\phi}\\
 sin(\theta/2)
\\
- cos(\theta/2) e^{-i\phi} 	
\end{bmatrix}
\end{equation} 
so that they become eigenvectors of the chiral projectors $P_{\pm}=(1\pm \gamma_5)/2$. 

Having specified these spinors, one can define the matrices
rotating one set to the another. In particular, the transition 
between the spin and helicity states,  takes the  simple form 

\begin{eqnarray} \label{eqn_spin_to_hel}
	{\langle s \uparrow | h+\rangle  \over \langle  h+ |  h+ \rangle }&=&cos(\theta/2)  ,\,\,\,
		{\langle s \downarrow | h+\rangle  \over \langle  h+ |  h+ \rangle }=sin(\theta/2)e^{i\phi} \nonumber\\
			{\langle s \uparrow | h- \rangle  \over \langle  h+ |  h+ \rangle }& =& -sin(\theta/2) ,\,\,\,
				{\langle s \downarrow | h-\rangle  \over \langle  h+ |  h+ \rangle }=cos(\theta/2)e^{i\phi} 
				\nonumber\\
\end{eqnarray}	
which -- in the ultrarelativistic limit -- is the same as 
the matrix between the spin-basis and chirality-basis.	

\section{LF Dirac spinors}~\label{APP-LFX}
The LF Dirac spinors used to derive (\ref{UVX1}) are for the L-quark spinor
with mass $m_{Q_1}$
\bea
U_L(k,\uparrow)&=&\frac 1{(\sqrt{2} k^+)^{\frac 12}}
\begin{pmatrix}
m_{Q_1}\\
-k_R
\end{pmatrix}\nonumber\\
U_L(k,\downarrow)&=&\frac 1{(\sqrt{2} k^+)^{\frac 12}}
\begin{pmatrix}
-k_L\\
\sqrt{2} k^++\frac 12 m_{Q_1}
\end{pmatrix}\nonumber\\
\eea
and the R-quark spinor with the same mass
\bea
U_R(k,\uparrow)&=&\frac 1{(\sqrt{2} k^+)^{\frac 12}}
\begin{pmatrix}
\sqrt{2} k^++ \frac 12 m_{Q_1}\\
k_R
\end{pmatrix}\nonumber\\
U_R(k,\downarrow)&=&\frac 1{(\sqrt{2} k^+)^{\frac 12}}
\begin{pmatrix}
k_L\\
 m_{Q_1}
\end{pmatrix}\nonumber\\
\eea
For the L-antiquark spinor with mass $m_{Q_2}$, we have
\bea
V_L(k,\uparrow)&=&\frac 1{(\sqrt{2} k^+)^{\frac 12}}
\begin{pmatrix}
-k_L\\
\sqrt{2} k^++ \frac 12 m_{Q_2}
\end{pmatrix}\nonumber\\
V_L(k,\downarrow)&=&\frac 1{(\sqrt{2} k^+)^{\frac 12}}
\begin{pmatrix}
 -m_{Q_2}\\
 k_R
\end{pmatrix}\nonumber\\
\eea
and for the R-antiquark with the same mass 
\bea
V_R(k,\uparrow)&=&\frac 1{(\sqrt{2} k^+)^{\frac 12}}
\begin{pmatrix}
-k_L\\
-m_{Q_2}
\end{pmatrix}\nonumber\\
V_R(k,\downarrow)&=&\frac 1{(\sqrt{2} k^+)^{\frac 12}}
\begin{pmatrix}
\sqrt{2}k^++\frac 12 m_{Q_2}\\
k_R
\end{pmatrix}\nonumber\\
\eea

\bibliography{allbib,string,meson1X,meson3}

\end{document}